\documentclass[3p,review]{elsarticle} 

\usepackage{amssymb}  
\usepackage{amsthm}   
\usepackage{lineno}   
\usepackage{amsmath}  
\usepackage{cases}    
\usepackage{tabularx} 
\usepackage[figuresright]{rotating}  
\usepackage{color}
\usepackage{float}
\usepackage{multirow}
\usepackage[hidelinks]{hyperref} 
\hypersetup{pdfstartview={XYZ null null 1.00}} 
\hypersetup{pdfpagemode=UseNone} 

\allowdisplaybreaks  

\journal{}
\date{March 24, 2021}

\begin{document}

\begin{frontmatter}

\title{High-Order Implicit Large-Eddy Simulation of Flow over a Marine Propeller}

\author[gwu,clarkson]{Bin Zhang \corref{cor1}}
\ead{bzh@gwmail.gwu.edu}

\author[clarkson]{Chi Ding}

\author[clarkson]{Chunlei Liang}

\cortext[cor1]{Corresponding author.}

\address[gwu]{Department of Mechanical and Aerospace Engineering, The George Washington University, Washington, DC 20052 USA}
\address[clarkson]{Department of Mechanical and Aeronautical Engineering, Clarkson University, Potsdam, NY 13676 USA}

\begin{abstract}
We report the first high-order eddy-resolving simulation of flow over a marine propeller using a recently developed high-order sliding-mesh method. This method employs the flux reconstruction framework and a new dynamic curved mortar approach to handle the complex rotating geometries. For a wide range of working conditions, it is validated to predict the loads very accurately against experiments. The method's low-dissipation characteristic has allowed the capturing of a broad spectrum of turbulence structures for very long distances even on a very coarse grid. Comparison with a previous low-order simulation is also carried out to show the low-dissipation advantage of the present simulations. From detailed load analysis, the major loads and their distributions and time and frequency scales are identified. Visualizations of the instantaneous, phase-averaged, and time-averaged flow fields have revealed the processes of tip vortex formation, major vortex evolutions, and flow instability developments at different working conditions. The effects of different fairwaters on the propeller's overall performance are also quantitatively assessed.
\end{abstract}

\begin{keyword}
high-order method \sep flux reconstruction \sep sliding mesh \sep implicit LES \sep marine propeller
\end{keyword}

\end{frontmatter}


\section{Introduction}
\label{sec:introduction}

Numerical techniques for studying marine propellers have seen immense advances in the past decades. For example, the lifting surface procedures based on vortex lattice and panel methods have been widely used in propeller design. A representative program in this category is the PSF code developed at MIT in the 1980's \cite{greeley-1982}. The vortex lattice and panel methods have very quick turnaround time, and can predict propeller loads rather accurately when combined with empirical vortex and wake models. They are, however, incapable of revealing the fine details of a flow field for in-depth analysis of flow physics such as flow instability and acoustics. Since the 1990's, the Reynolds-averaged Navier–Stokes (RANS) methods have gained popularity in simulating propeller flows. The RANS methods do provide richer flow field information than the inviscid vortex lattice and panel methods. But the averaging nature of a RANS (even an unsteady RANS) method still smooths out a lot of flow details, especially the small and intermediate instantaneous eddies. In more recent years, the increasing high-performance computing power has allowed eddy-resolving techniques, such as detached-eddy simulation (DES) and large-eddy simulation (LES), to have been performed on marine propellers in the scale of millions or even trillions of grid elements. For instance, \citet{muscari-2013} and \citet{di-2014} employed DES to study the vortex dynamics of a propeller in different flow conditions. \citet{verma-2012} and \citet{jang-2013} used LES to investigate the flow around a reverse rotating marine propeller. \citet{balaras-2015} applied an immersed boundary based LES technique to explore the flow around a propeller with and without an upstream appendage. \citet{kumar-2017} systematically studied the wake instability of a propeller using LES. However, all the aforementioned DES and LES approaches are of low-order accuracies (at most second-order) that could potentially introduce large numerical dissipation and dispersion to a flow field.

On the other hand, high-order (third and above) methods, especially polynomial based high-order methods, are seeing rapid development in recent years. High-order methods show many advantages over the traditional low-order ones. For example, for the same number of degrees of freedom, a high-order method produces solutions with much smaller numerical error than a low-order method does. Furthermore, a high-order method can employ high-order unstructured curved meshes that approximate curved geometries much more accurately than the linear meshes for a low-order method. The most popular high-order methods include the discontinuous Galerkin (DG) method \cite{reed-1973, cockburn-2011}, the spectral element method \cite{patera-1984, karniadakis-2005, kopriva-2009}, the spectral volume method \cite{wang-2002a, wang-2002b}, the spectral difference (SD) method \cite{kopriva-1996a, kopriva-1998, liu-2006, wang-2007a,jameson-2010,balan-2012}, etc. Among these methods, the SD method solves equations in differential form directly, and is one of the most efficient high-order methods. Recently, the ideas of collocating solution and flux points of the SD method and correcting fluxes using higher-degree polynomials have led to an even more efficient high-order method --- the flux reconstruction (FR) method \cite{huynh-2007, huynh-2009}, also known as the correction procedure via reconstruction (CPR) \cite{wang-2009}. Besides its better efficiency, by choosing different correction polynomials, the FR method can recover many existing high-order schemes such as DG and SD, and can even produce new schemes that had never been reported before. The stability of the FR method has been proved in \cite{jameson-2012}. The most recent developments on the FR method are summarized in \cite{wang-2016}.

To the authors' knowledge, there is still no reported high-order eddy-resolving simulation of marine propellers. One important reason is the severe challenge on how to treat the complex rotational geometry of a propeller in a high-order method without deteriorating the method's accuracies in both space and time. To tackle this challenge, \citet{zhang-2015a, zhang-2015b} developed a high-order sliding-mesh method for the SD and FR methods by introducing the concept of curved dynamic mortar elements. A parallelization approach was proposed for this method in \cite{zhang-2016a}, and the extension to three-dimensions was achieved in \cite{zhang-2016c}. More recently, this method was further extended to sliding interfaces with general nonuniform meshes \cite{zhang-2018}. An updated version that is high-order in time, arbitrarily high-order in space, provably conservative, and provably outflow preservative has also been established \cite{zhang-2020a}. In the present work, we apply this method to implicit LES (without using an explicit sub-grid-scale model) \cite{parsani-2010, uranga-2011, beck-2014, bull-2015, vermeire-2016} of the loads and flow fields of a real marine propeller at various working conditions. This is the first time a high-order method being applied to simulate a marine propeller.

The rest of this paper is organized as follows. Section \ref{sec:methods} gives a brief description of the flow equations and the numerical methods. Section \ref{sec:setup} details the simulation setup. Simulation results and discussions are reported in Section \ref{sec:results}. Finally, Section \ref{sec:summary} concludes this paper.

\section{Numerical methods}
\label{sec:methods}

\subsection{The physical equations}
We numerically solve the following three-dimensional unsteady Navier-Stokes equations in a conservative form,
\begin{equation}
\frac{\partial \mathbf{Q}} {\partial t} +  \frac{\partial \mathbf{F}} {\partial x} + \frac{\partial \mathbf{G}} {\partial y} + \frac{\partial \mathbf{H}} {\partial z} = \boldsymbol{0},
\label{eq:physical}
\end{equation}
where $\mathbf{Q}$ is the vector of conservative variables, $\mathbf{F}$, $\mathbf{G}$ and  $\mathbf{H}$ are the flux vectors in each coordinate direction. These terms have the following expressions,
\begin{align}
\mathbf{Q} &= [\rho  ~ \rho u  ~ \rho v  ~ \rho w  ~ E]^\mathsf{T}, \label{eq:Q} \\
\mathbf{F} &= \mathbf{F}_{\text{inv}}(\mathbf{Q}) + \mathbf{F}_{\text{vis}}(\mathbf{Q},\nabla \mathbf{Q}), \label{eq:F}\\
\mathbf{G} &= \mathbf{G}_{\text{inv}}(\mathbf{Q}) + \mathbf{G}_{\text{vis}}(\mathbf{Q},\nabla \mathbf{Q}), \label{eq:G}\\
\mathbf{H} &= \mathbf{H}_{\text{inv}}(\mathbf{Q}) + \mathbf{H}_{\text{vis}}(\mathbf{Q},\nabla \mathbf{Q}), \label{eq:H}
\end{align}
where $\rho$ is fluid density, $u$, $v$ and $w$ are the velocity components, $E$ is the total energy per volume defined as $E = p/(\gamma-1) + \frac{1}{2}\rho(u^2+v^2+w^2)$, $p$ is pressure, $\gamma$ is the ratio of specific heats and is set to 1.4.

The fluxes have been split into inviscid and viscous parts. The inviscid fluxes are only functions of the conservative variables and have the following expressions,
\begin{equation}
\mathbf{F}_{\text{inv}} = \left[
\begin{array}{c}
\rho u       \\
\rho u^2 + p \\
\rho uv      \\
\rho uw      \\
u(E+p)
\end{array}
\right], \
\mathbf{G}_{\text{inv}} = \left[
\begin{array}{c}
\rho v       \\
\rho uv      \\
\rho v^2 + p \\
\rho vw      \\
v(E+p)
\end{array}
\right],\
\mathbf{H}_{\text{inv}} = \left[
\begin{array}{c}
\rho w       \\
\rho uw      \\
\rho vw      \\
\rho w^2 + p \\
w(E+p)
\end{array}
\right].
\label{eq:FGHinv}
\end{equation}
The viscous fluxes are functions of the conservative variables and their gradients. The expressions are
\begin{equation}
\mathbf{F}_{\text{vis}} = -\left[
\begin{array}{c}
0         \\
\tau_{xx} \\
\tau_{yx} \\
\tau_{zx} \\
u\tau_{xx}+v\tau_{yx}+w\tau_{zx}+ \kappa T_x
\end{array}
\right],
\end{equation}
\begin{equation}
\mathbf{G}_{\text{vis}} = -\left[
\begin{array}{c}
0         \\
\tau_{xy} \\
\tau_{yy} \\
\tau_{zy} \\
u\tau_{xy}+v\tau_{yy}+w\tau_{zy}+\kappa T_y
\end{array}
\right],
\end{equation}
\begin{equation}
\mathbf{H}_{\text{vis}} = -\left[
\begin{array}{c}
0         \\
\tau_{xz} \\
\tau_{yz} \\
\tau_{zz} \\
u\tau_{xz}+v\tau_{yz}+w\tau_{zz}+\kappa T_y
\end{array}
\right],
\end{equation}
where $\tau_{ij}$ is shear stress tensor which is related to velocity gradients as $\tau_{ij} = \mu (u_{i,j}+u_{j,i}) + \lambda\delta_{ij}u_{k,k}$, $\mu$ is dynamic viscosity, $\lambda=-2/3\mu$ based on Stokes' hypothesis, $\delta_{ij}$ is the Kronecker delta, $\kappa$ is thermal conductivity, $T$ is temperature that is related to density and pressure through the ideal gas law $p=\rho \mathcal{R} T$, where $\mathcal{R}$ is the gas constant.

\subsection{The computational equations}
As will be discussed later, we map each moving grid element from the physical space to a stationary standard cubic element in the computational space where the equations are solved. Assume the mapping is: $t=\tau$, $x=x(\tau,\xi,\eta,\zeta)$, $y=y(\tau,\xi,\eta,\zeta)$ and $z=z(\tau,\xi,\eta,\zeta)$, where $(\tau,\xi,\eta,\zeta)$ are the computational time and space. It can be shown that the flow equations will take the following conservative form in the computational space,
\begin{equation}
\frac{\partial \widetilde{\mathbf{Q}}} {\partial t} +  \frac{\partial \widetilde{\mathbf{F}}} {\partial \xi} + \frac{\partial \widetilde{\mathbf{G}}} {\partial \eta} + \frac{\partial \widetilde{\mathbf{H}}} {\partial \zeta} = \boldsymbol{0},
\label{eq:computational}
\end{equation}
and the computational variable and fluxes are related to the physical ones through
\begin{equation}
\left[
\begin{array}{c}
\widetilde{\mathbf{Q}} \\
\widetilde{\mathbf{F}} \\
\widetilde{\mathbf{G}} \\
\widetilde{\mathbf{H}}
\end{array}
\right]
= |\mathcal{J}|\mathcal{J}^{-1}
\left[
\begin{array}{c}
\mathbf{Q} \vphantom{\widetilde{\mathbf{Q}}} \\
\mathbf{F} \vphantom{\widetilde{\mathbf{F}}} \\
\mathbf{G} \vphantom{\widetilde{\mathbf{G}}} \\
\mathbf{H} \vphantom{\widetilde{\mathbf{H}}}
\end{array}
\right],
\label{eq:transform_relation}
\end{equation}
where $|\mathcal{J}|$ is determinant of the Jacobian matrix and $\mathcal{J}^{-1}$ is the inverse Jacobian matrix, and their expressions are
\begin{gather}
|\mathcal{J}| = \left| \frac{\partial(t, x,y,z)}{\partial(\tau, \xi,\eta,\zeta)} \right| = \left|
\begin{array}{cccc}
1   & 0       & 0        & 0         \\
x_t & x_{\xi} & x_{\eta} & x_{\zeta} \\
y_t & y_{\xi} & y_{\eta} & y_{\zeta} \\
z_t & z_{\xi} & z_{\eta} & z_{\zeta}
\end{array} \right|, \\
\mathcal{J}^{-1} = \frac{\partial(\tau, \xi, \eta, \zeta)}{\partial(t,x,y,z)} = \left[
\begin{array}{cccc}
1       & 0       & 0       & 0       \\
\xi_t   & \xi_x   & \xi_y   & \xi_z   \\
\eta_t  & \eta_x  & \eta_y  & \eta_z  \\
\zeta_t & \zeta_x & \zeta_y & \zeta_z \\
\end{array}
\right].
\end{gather}

Besides the flow equations, the Geometric Conservation Law (GCL) \cite{thomas-1979} also needs to be numerically satisfied to ensure free-stream preservation on moving grids. The GCL equations and the steps for solving them are described in our previous papers \cite{zhang-2015b,zhang-2016c}.

\subsection{The flux reconstruction method}
The meshes in this work consist of hexahedral elements only. The first step of the FR method is to map each hexahedral element to a standard cubic element of unit size, i.e., $0\leq \xi, \eta, \zeta \leq 1$. This can be done using the following iso-parametric mapping,
\begin{equation}
\left[
\begin{array}{c}
x \\ y \\ z
\end{array} \right] =
\sum_{i=1}^{K} M_i(\xi,\eta,\zeta)
\left[
\begin{array}{c}
x_i(t) \\ y_i(t) \\ z_i(t)
\end{array} \right],
\label{eq:mapping}
\end{equation}
where $K$ is the number of nodes of a hexahedral element, $M_i$ (detailed expressions can be found in, for example, \cite{bathe-2006}) is the shape function of the $i$-th node, and $(x_i,y_i,z_i)$ are the coordinates of the $i$-th node.

\begin{figure}[H]
\centering
\includegraphics[width=2in]{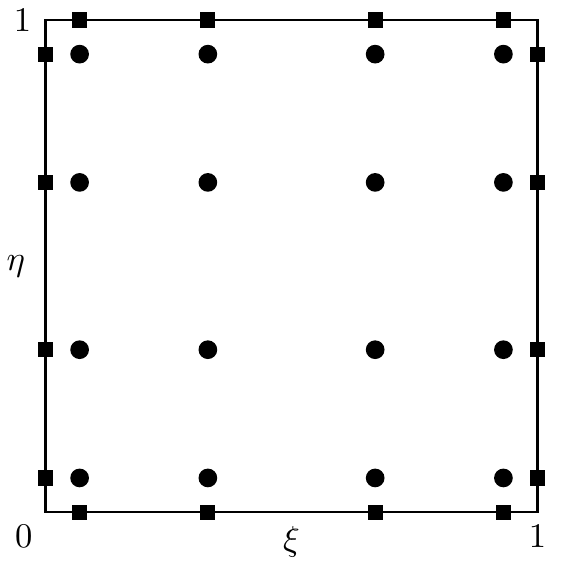}
\caption{Schematic of solution points (circles) and flux points (squares) in the $\xi\textnormal{-}\eta$ plane for a fourth-order FR method.}
\label{fig:spfp}
\end{figure}

Next, solution points (SPs, denoted by $X_s$) and flux points (FPs, denoted by $X_f$) are defined along each coordinate direction in the standard element. Fig. \ref{fig:spfp} shows a schematic of the distribution of the SPs and FPs in the $\xi$-$\eta$ plane for a fourth-order FR method. Generally, for an $N$-th order FR scheme, there are $N$ SPs and FPs in each direction, where the SPs are in the interior and the FPs are on the boundaries of the standard element. The SPs and FPs are chosen as the Legendre points in this work.

Subsequently, Lagrange interpolation bases are defined at each SP. For example, at the $i$-th SP we have
\begin{equation}
h_i(X) = \prod_{s=1,s\neq i}^{N}\left(\frac{X-X_s}{X_i-X_s}\right).
\label{eq:hbasis}
\end{equation}
The resulting bases also form a basis of polynomials of degrees less than or equal to $N-1$, i.e., $\boldsymbol{\mathsf{P}}_{N\!-\!1}$. These interpolation bases allow the construction of solution and flux polynomials inside each element through tensor products. For example,
\begin{align}
\widetilde{\mathbf{Q}}(\xi,\eta,\zeta) &= \sum_{k=1}^{N} \sum_{j=1}^{N} \sum_{i=1}^{N} \widetilde{\mathbf{Q}}_{ijk} h_i(\xi) h_j(\eta) h_k(\zeta), \label{eq:Qt} \\
\widetilde{\mathbf{F}}(\xi,\eta,\zeta) &= \sum_{k=1}^{N} \sum_{j=1}^{N} \sum_{i=1}^{N} \widetilde{\mathbf{F}}_{ijk} h_i(\xi) h_j(\eta) h_k(\zeta), \label{eq:Ft}
\end{align}
where the subscript $ijk$ denotes the discrete value at the $ijk$-th SP. All these polynomials are in $\boldsymbol{\mathsf{P}}_{N\!-\!1,N\!-\!1,N\!-\!1}$.

The above solution and flux polynomials are continuous within each element, but discontinuous across cell boundaries. Therefore, common values need to be defined on cell boundaries. There are various ways to calculate these common values. In this work, the common solution is calculated as the average of the discontinuous values from the two sides of a boundary; the common inviscid fluxes are computed using the Rusanov Riemann solver\cite{rusanov-1961}; the common viscous fluxes are computed from the common solutions and common gradients.

There is one more issue: after taking the first-order spatial derivatives in the governing equations, the three flux terms become elements in $\boldsymbol{\mathsf{P}}_{N\!-\!2,N\!-\!1,N\!-\!1}$, $\boldsymbol{\mathsf{P}}_{N\!-\!1,N\!-\!2,N\!-\!1}$, and $\boldsymbol{\mathsf{P}}_{N\!-\!1,N\!-\!1,N\!-\!2}$, respectively, and are inconsistent with the solution term. To fix this issue, the original fluxes need to be reconstructed. This is done by using correction functions that are polynomials of degree no less than $N$. Taking the flux in the $\xi$ direction as an example, the reconstructed flux polynomials is
\begin{equation}
\widehat{\mathbf{F}}(\xi,\eta,\zeta) = \widetilde{\mathbf{F}}(\xi,\eta,\zeta) + \big[\widetilde{\mathbf{F}}^\text{com}(0,\eta,\zeta) - \widetilde{\mathbf{F}}(0,\eta,\zeta)\big] \cdot g_\text{\tiny L}(\xi) + [\widetilde{\mathbf{F}}^\text{com}(1,\eta,\zeta) - \widetilde{\mathbf{F}}(1,\eta,\zeta)] \cdot g_\text{\tiny R}(\xi)
\end{equation}
where $\widetilde{\mathbf{F}}(\xi,\eta,\zeta)$ is from (\ref{eq:Ft}); $\widetilde{\mathbf{F}}^{\text{com}}$ represents the common flux on a cell boundary; $g_\text{\tiny L}$ and $g_\text{\tiny R}$ are the left and right correction functions, and are required to at least satisfy
\begin{equation}
\begin{alignedat}{2}
g_\text{\tiny L}(0) &= 1, \quad g_\text{\tiny L}(1) &&= 0, \\
g_\text{\tiny R}(0) &= 0, \quad g_\text{\tiny R}(1) &&= 1,
\end{alignedat}
\end{equation}
which ensures that
\begin{equation}
\widehat{\mathbf{F}}(0,\eta,\zeta) = \widetilde{\mathbf{F}}^\text{com}(0,\eta,\zeta), \quad \widehat{\mathbf{F}}(1,\eta,\zeta) = \widetilde{\mathbf{F}}^\text{com}(1,\eta,\zeta),
\end{equation}
i.e., the reconstructed fluxes still take the common values at cell interfaces. In this work, we have employed the $g_{\text{\tiny DG}}$ correction function \cite{huynh-2007}. The other two fluxes are reconstructed in the same way.

Finally, the governing equations can be written in the following residual form,
\begin{equation}
\left. \frac{\partial \widetilde{\mathbf{Q}}} {\partial t} \right|_{ijk} = - \left[ \frac{\partial \widehat{\mathbf{F}}}{\partial\xi} +
\frac{\partial \widehat{\mathbf{G}}}{\partial\eta} + \frac{\partial \widehat{\mathbf{H}}}{\partial\zeta} \right]_{ijk} = \mathfrak{R}_{ijk}, \quad i,j,k=1,2,\cdots,N, \label{eq:resid}
\end{equation}
where $\mathfrak{R}_{ijk}$ is the residual at the $(i,j,k)$-th SP. This system can be time marched using either explicit or implicit temporal schemes.

\subsection{A sliding-mesh method}
\begin{figure}[H]
\centering
\includegraphics[width=1.6in]{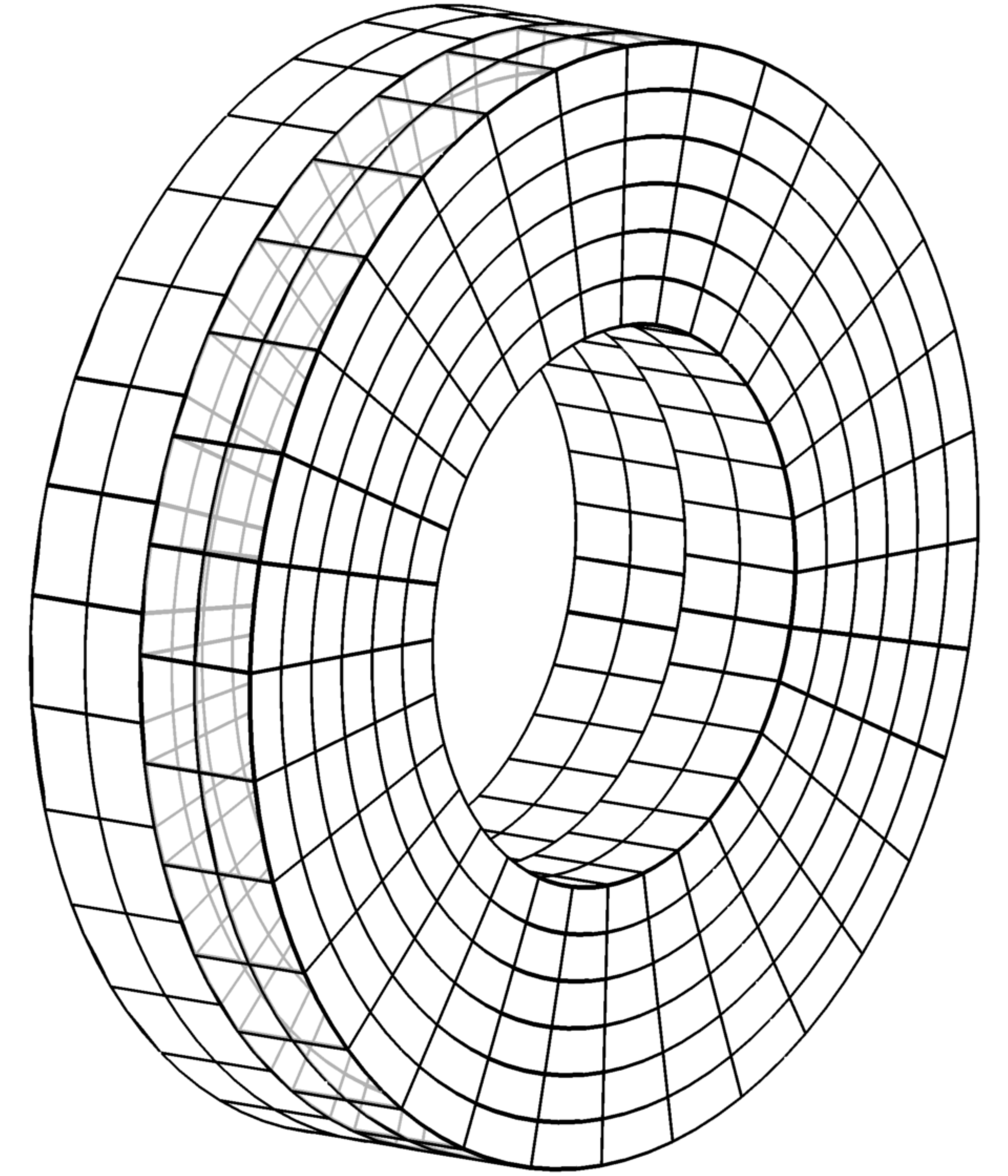} \qquad\qquad
\includegraphics[width=1.6in]{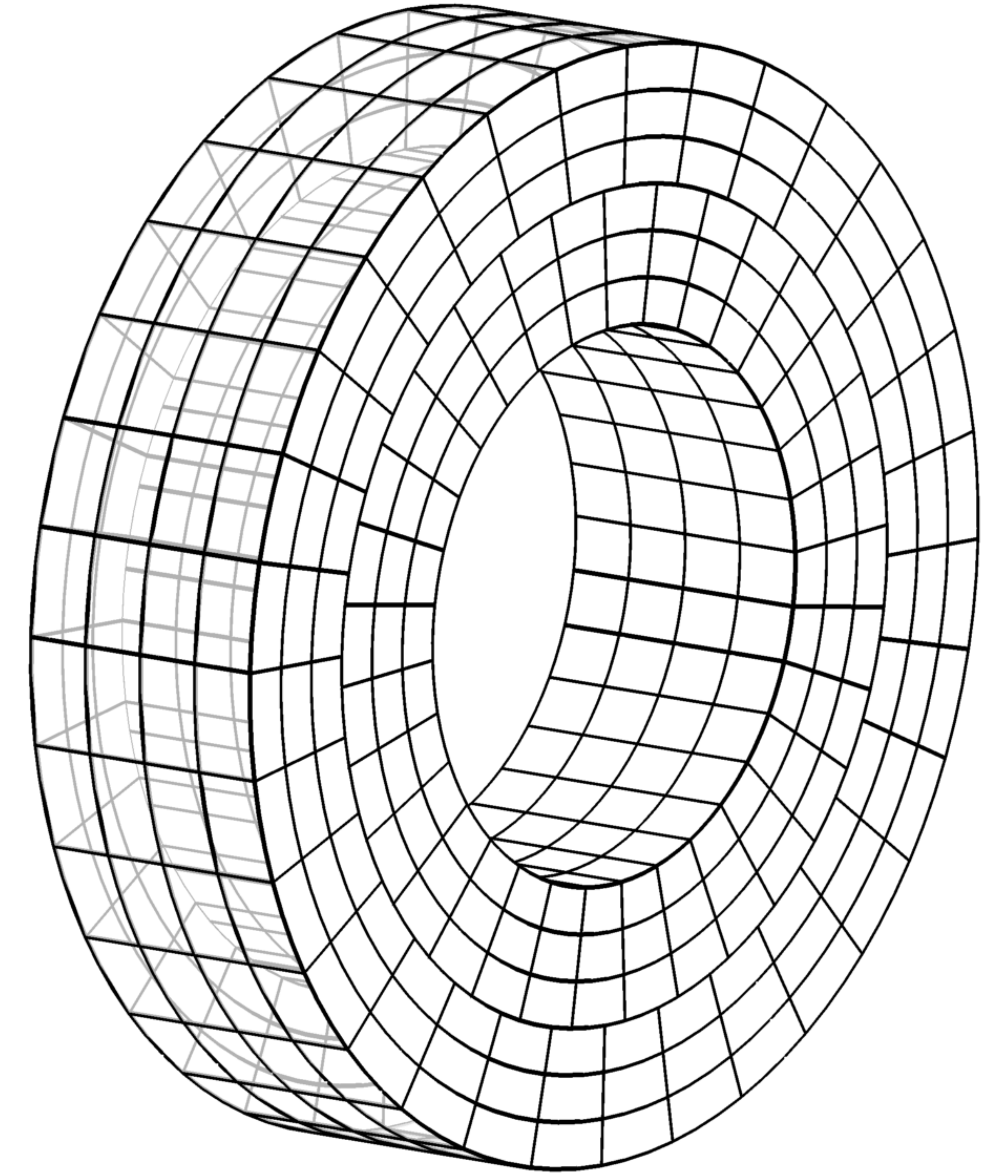}
\caption{Two types of sliding meshes: left, annular sliding; right, cylindrical sliding.}
\label{fig:sliding_mesh}
\end{figure}

In three-dimensions, we consider two types of sliding interfaces as depicted in Fig. \ref{fig:sliding_mesh}: one is annular, and the other is cylindrical. To simplify the explanation, we have required the mesh to only unmatch in the azimuthal direction but match in the radial (for annular sliding) or axial (for cylindrical sliding) direction. We also require equal mesh size in the azimuthal direction. These restrictions are imposed for explanation purposes only and can be easily lifted in practice. More details can be found in our previous papers \cite{zhang-2015a,zhang-2015b,zhang-2016a,zhang-2016c,zhang-2018,zhang-2020a}.

\begin{figure}[H]
\centering
\includegraphics[width=4.5in]{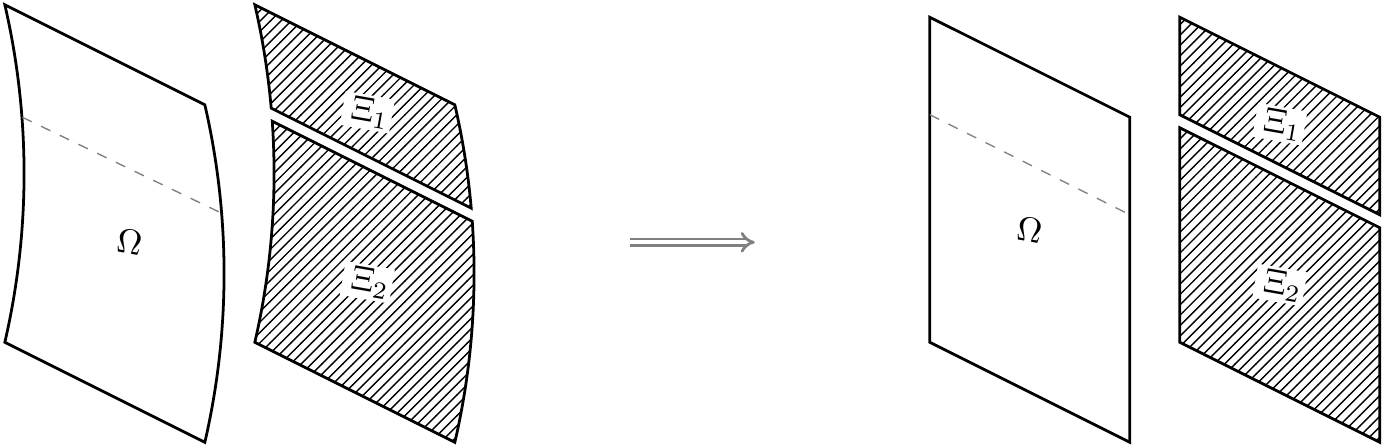}
\caption{Map curved cell face and mortars to straight ones.}
\label{fig:mortar_map}
\end{figure}

Since the SPs on the two sides of a nonconforming sliding interface do not match, we aim to find the best-possible common values of a variable on the two sides of a sliding interface. This can be achieved by least-squares projections using mortar elements as the intermediate medium. A mortar element is formed by the overlapping region of two cell faces. Taking the cylindrical sliding interface as an example, based on the assumptions we have made, a cell face has two mortar elements as sketched in Fig. \ref{fig:mortar_map}. The first step is to map a cell face and the mortars to standard ones as shown in the same figure, using, for example, iso-parametric mapping or transfinite mapping.

Let the $(\xi',\eta')$ denote the mortar space, then the computational and the mortar spaces are related as
\begin{equation}
\xi  =o + s \xi', ~~~ \eta = \eta',
\label{eq:xz}
\end{equation}
where $0\le \xi, \eta, \xi', \eta' \le 1$, and $o$ and $s$ are the offset and scaling of a mortar with respect to a cell face.

\begin{figure}[H]
\centering
\includegraphics[width=4.5in]{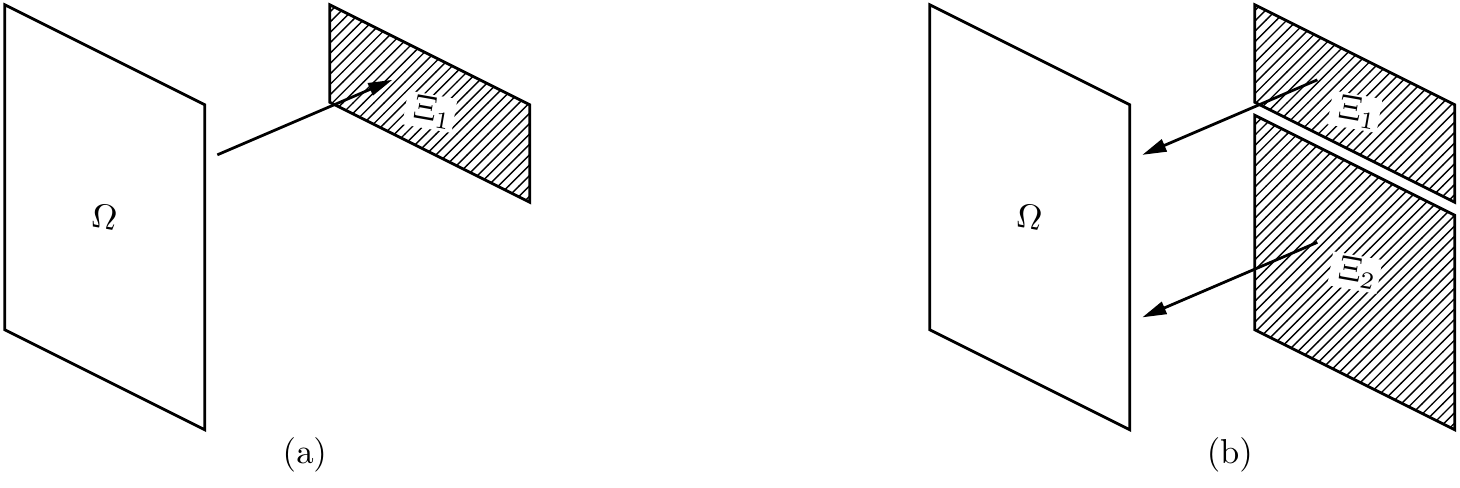}
\caption{Projection between face and mortar: (a) from face to the left side of a mortar, (b) from two mortars back to a face.}
\label{fig:mortar_prj}
\end{figure}

Let $\phi$ represent the variable of interest, and obviously it can be represented by the following polynomials on a cell face $\Omega$ and on the left side of a mortar $\Xi$,
\begin{align}
\phi^{\Omega}(\xi,\eta) &= \sum_{j=1}^{N} \sum_{i=1}^{N} \phi_{ij}^{\Omega} h_i(\xi) h_j(\eta), \label{eq:phi_face} \\
\phi^{\Xi,L}(\xi',\eta') &= \sum_{j=1}^{N} \sum_{i=1}^{N} \phi_{ij}^{\Xi,L} h_i(\xi') h_j(\eta'), \label{eq:phi_mortar}
\end{align}
where $\phi_{ij}^{\Omega}$ and $\phi_{ij}^{\Xi,L}$ are the discrete values at the $(i,j)$-th SP on $\Omega$ and the left side of $\Xi$, respectively. The $(\phi_{ij}^{\Xi,L})$'s are unknown, and can be obtained through the following projection (refer to Fig. \ref{fig:mortar_prj}(a)),
\begin{equation}
\int_0^1 \int_0^1 (\phi^{\Xi,L}(\xi',\eta') - \phi^{\Omega}(\xi,\eta)) h_\alpha(\xi')h_\beta(\eta') \text{d}\xi' \text{d}\eta' = 0,~ \forall\, \alpha,\beta=1,2,...,N.
\label{eq:prj1a}
\end{equation}
Considering the relations in (\ref{eq:xz}), it can be shown that the above two-dimensional projection is equivalent to the following one-dimensional one,
\begin{equation}
\int_0^1 (\phi^{\Xi,L}(\xi',X_j) - \phi^{\Omega}(\xi,X_j)) h_\alpha(\xi') \text{d} \xi' = 0,~ \forall\, \alpha=1,2,...,N,
\label{eq:prj1b}
\end{equation}
where $X_j$ is the coordinate of the $j$-th SP. Evaluating the above equation for all the $\alpha$'s, we will get a system of equations about $\phi_{1:N,j}^{\Xi,L}$. Repeating this process for every $j$, we will obtain every $\phi_{ij}^{\Xi,L}$. The values on the right side of a mortar, i.e., the $(\phi_{ij}^{\Xi,R})$'s, can be obtained in the same way.

After that, we are able to compute a common value on the mortar, e.g., through averaging or Riemann solver. Let us denote this common value as $\Phi^\Xi$. We then project this common variable back to a cell face from mortars as demonstrated in Fig. \ref{fig:mortar_prj}(b). And the projection is
\begin{equation}
\begin{split}
\int_{\xi=0}^{\xi=o_2} & \int_{\eta=0}^{\eta=1} (\Phi^{\Omega}(\xi,\eta) - \Phi^{\Xi_1}(\xi',\eta')) h_\alpha(\xi) h_\beta(\eta) \text{d}\xi \text{d}\eta   \\
&+ \int_{\xi=o_2}^{\xi=1} \int_{\eta=0}^{\eta=1} (\Phi^{\Omega}(\xi,\eta) - \Phi^{\Xi_2}(\xi',\eta')) h_\alpha(\xi) h_\beta(\eta) \text{d}\xi \text{d}\eta = 0, ~ \forall\, \alpha,\beta=1,2,...,N,
\label{eq:prj2a}
\end{split}
\end{equation}
where $\Phi^\Omega$ represents the polynomial of the unknown common variable on face $\Omega$. Similarly, this projection is equivalent to the following one-dimensional projection,
\begin{equation}
\int_{0}^{o_2} (\Phi^{\Omega}(\xi,X_j) - \Phi^{\Xi_1}(\xi',X_j)) h_\alpha(\xi) \text{d}\xi + \int_{o_2}^{1} (\Phi^{\Omega}(\xi,X_j) - \Phi^{\Xi_2}(\xi',X_j)) h_\alpha(\xi) \text{d}\xi = 0, ~ \forall\, \alpha =1,2,...,N,
\label{eq:prj2b}
\end{equation}
from which $\Phi_{1:N,j}^{\Omega}$ are obtained, and then also every $\Phi_{ij}^{\Omega}$ by repeating this process for $j$.

\section{Simulation setup}
\label{sec:setup}

The propeller studied in this work is the DTMB 4119 model designed at the David Taylor Model Basin \cite{denny-1968,jessup-1984,jessup-1989}. Its parameters are summarized in Tab. \ref{tab:propellerparam}, and the geometry is given in Tab. \ref{tab:propellergeo}, where $r$ represents radial position, $R=D/2$ is propeller radius, $c$ is blade section chord length, $P$ is section pitch, $\phi_P$ is section pitch angle, $t$ is section thickness, and $f$ is section camber.
\begin{table}[H]
\setlength{\tabcolsep}{2mm}
\centering
\begin{tabular}{>{\raggedright}m{5.0cm} >{\centering\arraybackslash}m{3cm}}
\hline
parameter                   &  value          \\
\hline
Number of blades $Z$        &   3             \\
Diameter $D$ [m]            & 0.305           \\
Hub diameter ratio $D_h/D$  &  $0.2$          \\
Design advance ratio $J$    & 0.833           \\
Rotation                    & Right handed    \\
Section thickness           & NACA66 modified \\
Section mean line           & NACA, $a=0.08$  \\
\hline
\end{tabular}
\caption{Design parameters of the DTMB 4119 propeller.}
\label{tab:propellerparam}
\end{table}
\begin{table}[H]
\setlength{\tabcolsep}{3mm}
\centering
\begin{tabular}{m{7mm} m{1cm} m{1cm} m{9mm} m{1.2cm} m{1.2cm}}
\hline
$r/R$ & ~~$c/D$ & $P/D$ & $\phi_P$ [$^\circ$] & ~~~$t/c$ & ~~~$f/c$ \\
\hline
0.2  & 0.3200 & 1.105 & 60.38 & 0.20550 & 0.01429 \\
0.3  & 0.3625 & 1.102 & 49.47 & 0.15530 & 0.02318 \\
0.4  & 0.4048 & 1.098 & 41.15 & 0.11800 & 0.02303 \\
0.5  & 0.4392 & 1.093 & 34.84 & 0.09016 & 0.02182 \\
0.6  & 0.4610 & 1.088 & 29.99 & 0.06960 & 0.02072 \\
0.7  & 0.4622 & 1.084 & 26.24 & 0.05418 & 0.02003 \\
0.8  & 0.4347 & 1.081 & 23.28 & 0.04206 & 0.01967 \\
0.9  & 0.3613 & 1.079 & 20.88 & 0.03321 & 0,01817 \\
0.95 & 0.2775 & 1.077 & 19.84 & 0.03228 & 0.01631 \\
1.0  & 0.0    & 1.075 & 18.89 & 0.03160 & 0.01175 \\
\hline
\end{tabular}
\caption{Geometry of the DTMB 4119 propeller.}
\label{tab:propellergeo}
\end{table}

When the geometry is given, a propeller flow is governed by two nondimensional parameters: advance ratio and Reynolds number. The advance ratio is defined as
\begin{equation}
J = \frac{U_\infty}{n D},
\end{equation}
where $U_\infty$ is the incoming flow speed, $n$ is propeller's revolution per second (RPS), and $D$ is propeller diameter. The following Reynolds number is usually adopted in experimental studies of marine propellers
\begin{equation}
Re_c = \frac{c_{0.7}\, U_{0.7}}{\nu} = \frac{c_{0.7} \sqrt{U_\infty^2 + (2\pi 0.7 R n)^2}}{\nu},
\end{equation}
where $c_{0.7}$ and $U_{0.7} = \sqrt{U_\infty^2 + (2\pi 0.7 R n)^2}$ are the chord length and the relative speed, respectively, at $r/R=0.7$, and $\nu$ is fluid kinematic viscosity. For numerical simulations, it is more convenient to define the Reynolds number as,
\begin{equation}
Re_D = \frac{D \, U_\infty}{\nu}.
\end{equation}
It can be shown that these two Reynolds numbers are conveniently convertible through the following relation,
\begin{equation}
Re_D = \frac{Re_c}{\frac{c_{0.7}}{D} \sqrt{1 + \left(\frac{0.7\pi}{J}\right)^2}}.
\end{equation}
Depending on $c_{0.7}/D$ and $J$, $Re_D$ could be either larger or smaller than $Re_c$, but usually not much different.

\begin{figure}[H]
\centering
\includegraphics[scale=0.9]{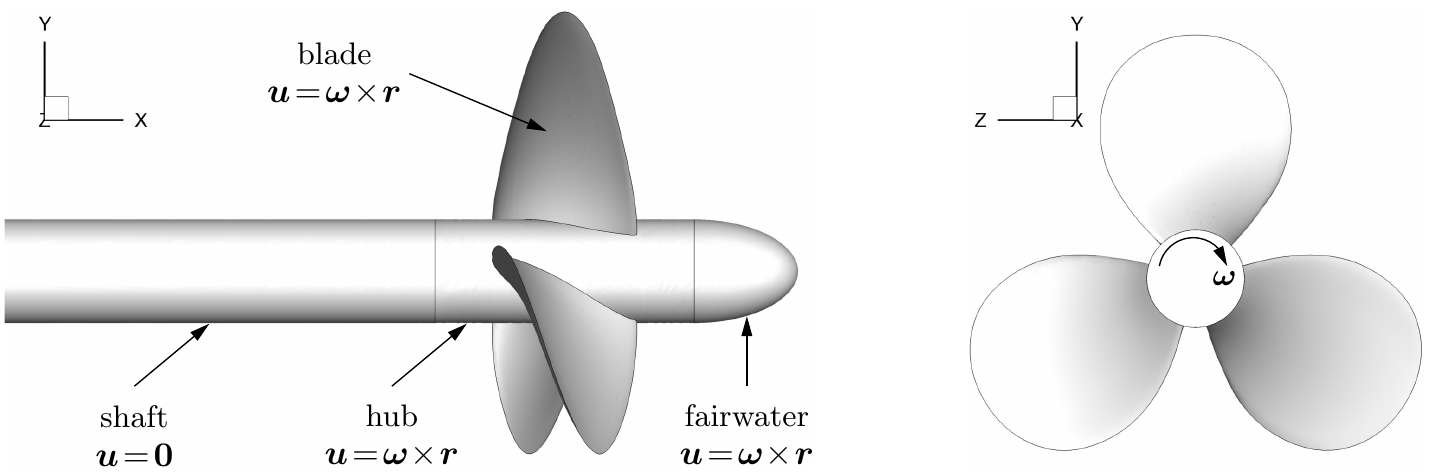}
\caption{Two views of the DTMB 4119 propeller: left, side view; right, back view (pressure side).}
\label{fig:geo_ellip}
\end{figure}

The geometry is visualized in Fig. \ref{fig:geo_ellip}. As a typical screw-type propeller, DTMB 4119 consists of four components: the shaft, the blades, the hub, and the fairwater. All components, except the shaft, rotate at an angular speed $\omega$. Hub and fairwater are usually not part of propeller design. In this work, the hub has a length $L_h=0.5D$, with the three blades installed evenly along the circumferential direction of the mid-hub. The fairwater in the figure is a $1:2$ ellipsoid, but cylindrical and hemispherical fairwaters are also employed in a later section to study their effects.

\begin{figure}[H]
\centering
\includegraphics[width=2.7in]{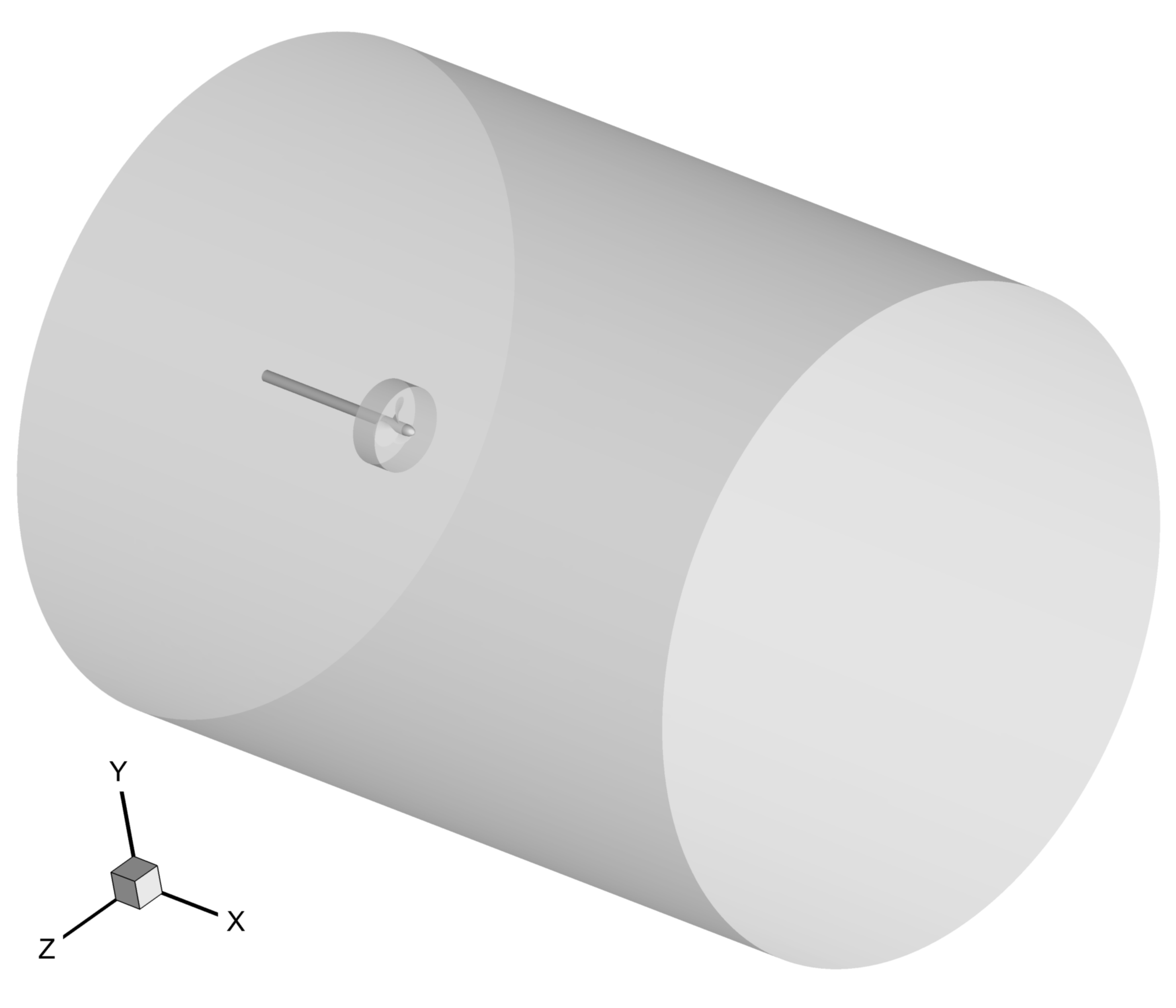} \qquad
\includegraphics[width=2.7in]{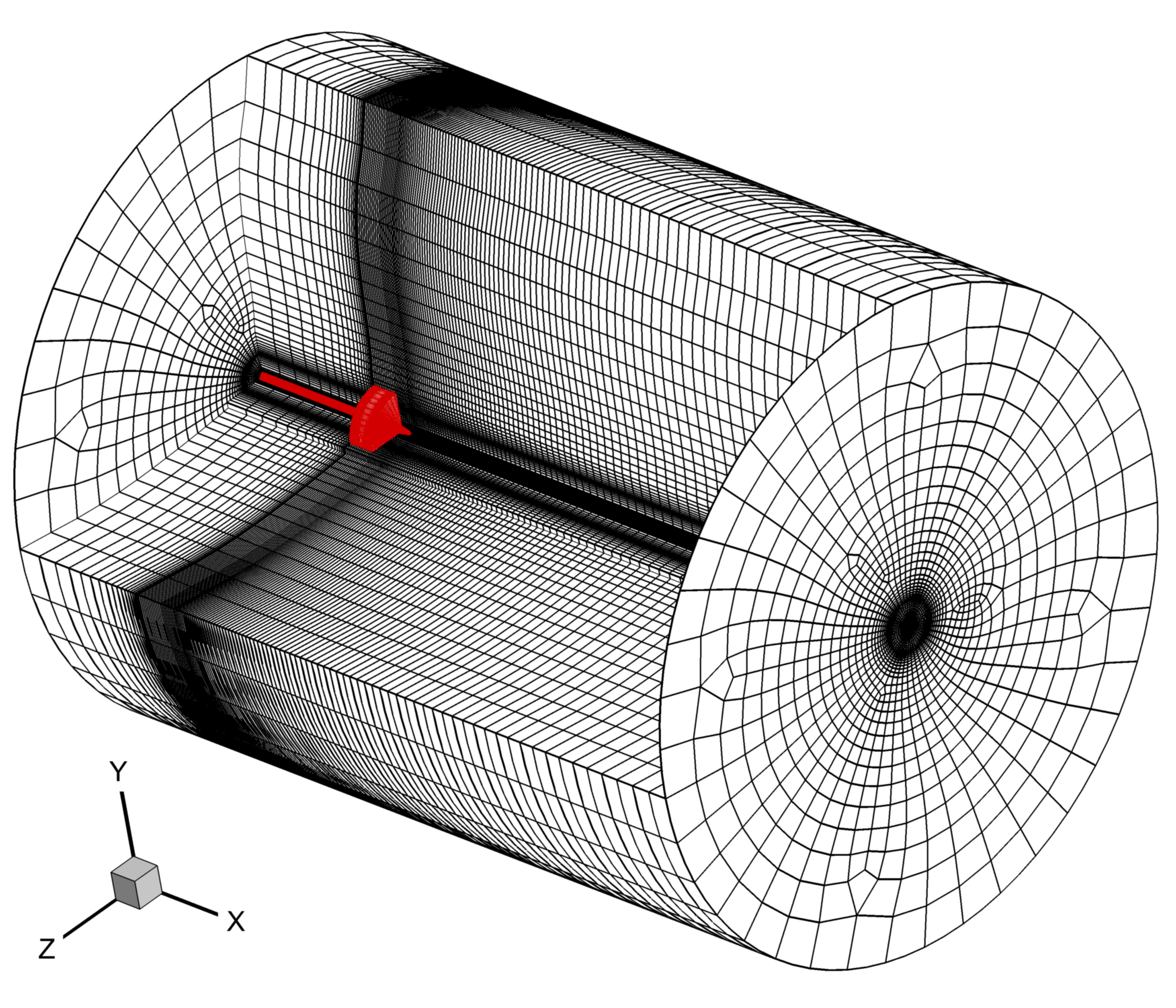}
\caption{Overall computational domain (left) and global view of the mesh (right).}
\label{fig:ellip_dom}
\end{figure}

\begin{figure}[H]
\centering
\includegraphics[width=2.5in]{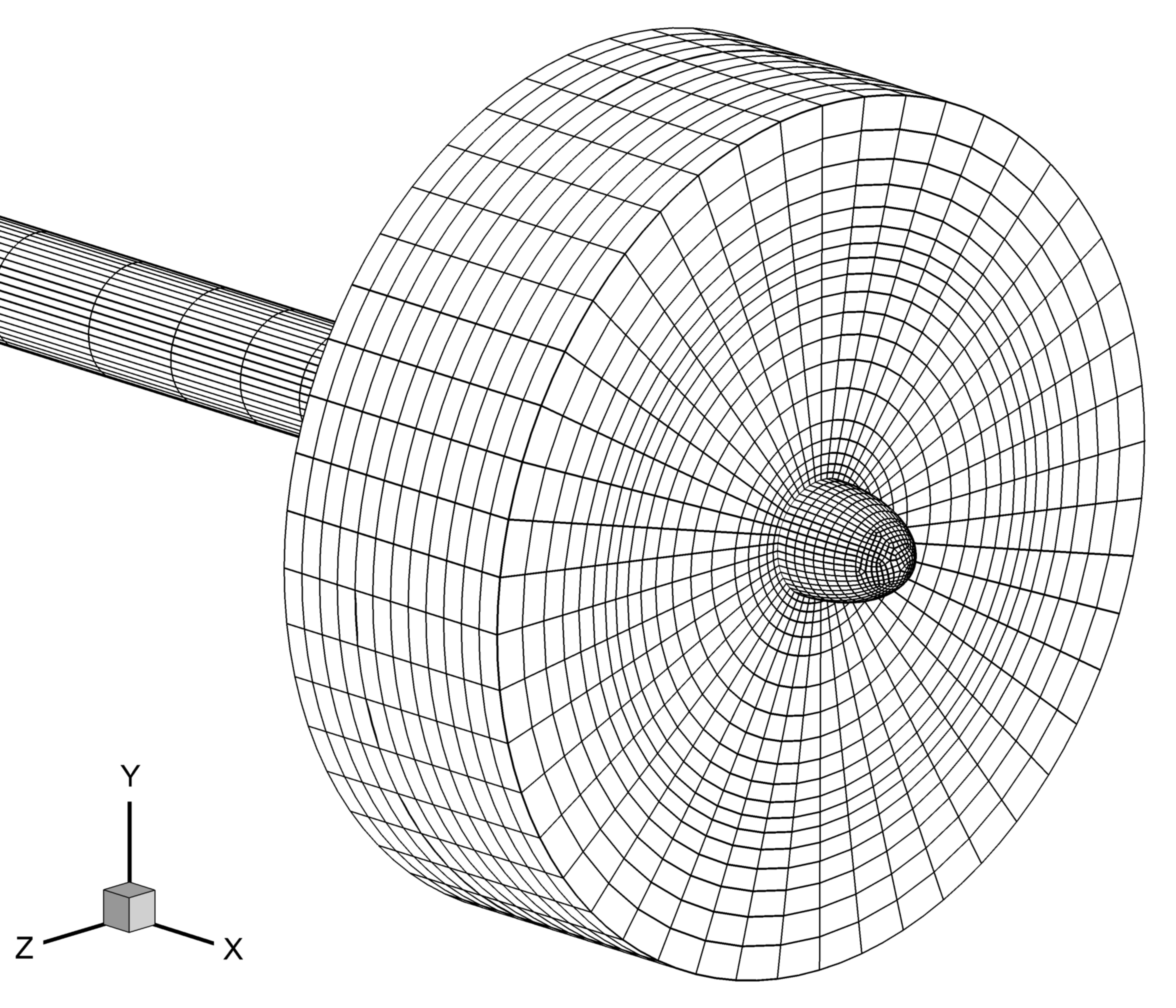} \qquad
\includegraphics[width=2.5in]{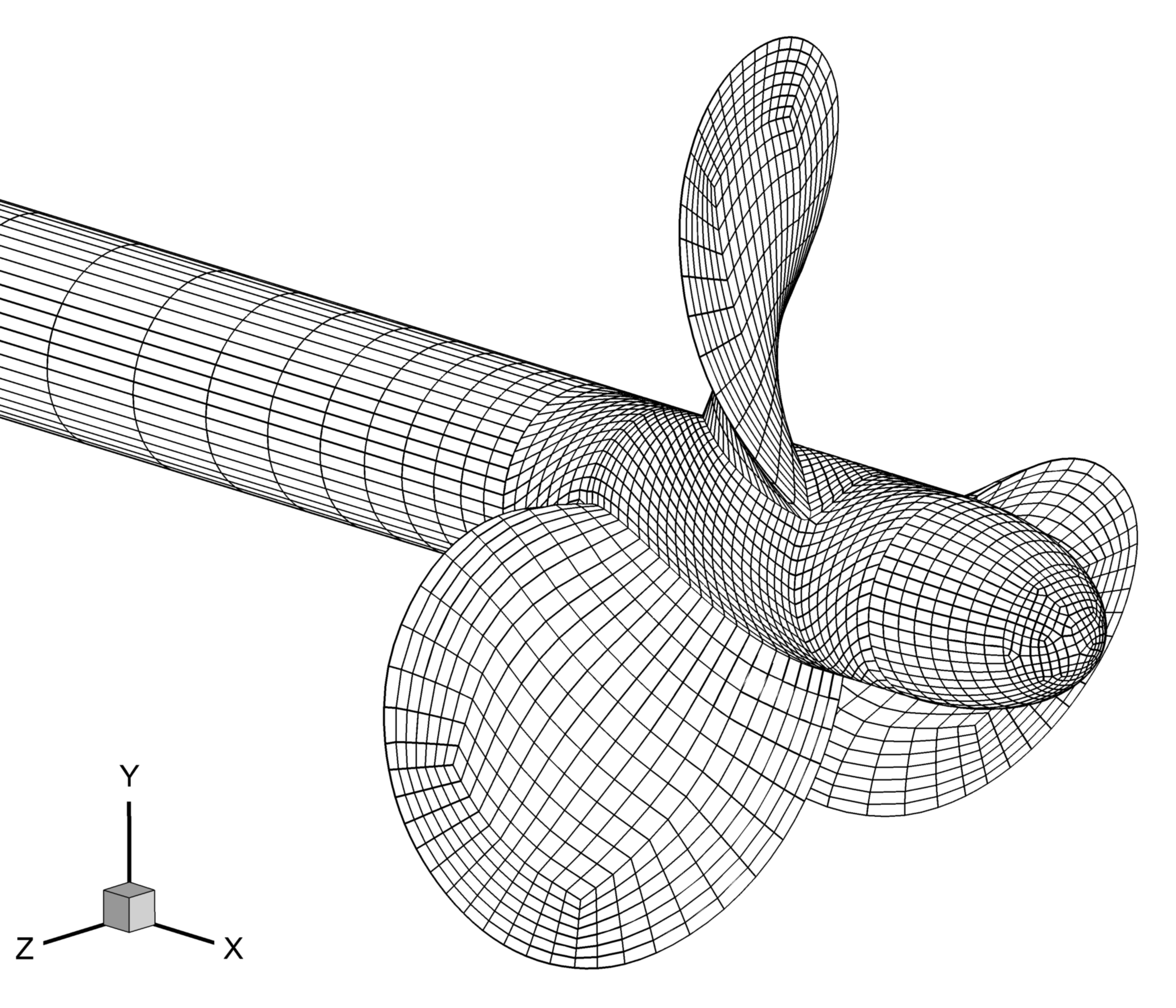}
\caption{Close views of the mesh on the sliding interfaces (left) and on the propeller surfaces (right).}
\label{fig:ellip_msh}
\end{figure}

The overall computational domain is cylindrical as shown in Fig. \ref{fig:ellip_dom}. It has a length of $15D$ in the streamwise (i.e., $x$) direction and a diameter of $12D$. The resulting blockage ratio of this domain with respect to the propeller is $0.69\%$, which is small enough to guarantee negligible confinement effects according to the study in \cite{wilson-1994}. The propeller locates $3D$ downstream from the inlet, with the shaft extends all the way to the inlet. The blades and the hub are enclosed in a small sliding disk region whose radius and thickness are $0.75D$ and $0.5D$, respectively. A global view of the mesh (with a $1/4$ cutout to expose the propeller) is shown on the right of Fig. \ref{fig:ellip_dom}, where the propeller and the sliding interfaces are colored in red.
The overall mesh consists of about $235,000$ quadratic curved hexahedral elements, of which about $36,000$ are within the small sliding disk region. The mesh is refined around the propeller as well as in the wake region. The first layer of the off wall elements on each blade surface has a height of approximately $0.015D$, and the first off wall solution point is about $0.0007D$ (for the fifth-order scheme) away from the walls. Two local views of the meshes on the sliding interfaces and the blade surfaces are shown in Fig. \ref{fig:ellip_msh}. We see that the high-order curved mesh captures the curvatures of the blade surfaces very well.

We treat the inlet as a Dirichlet boundary, the outer cylindrical surface and the outlet as characteristic farfields that allow waves and flow to leave without reflection \cite{jameson-1983}, and all solid surfaces as no-slip adiabatic walls. The incoming freestream flow has a low Mach number of $Ma_\infty=0.05$ so that compressibility effects are small. The Reynolds number is $Re_D=5.59\times 10^5$, which is equivalent to $Re_c=7.3\times 10^5$. Various advance coefficients are studied but with a focus on the design value (i.e., $J=0.833$). The nondimensional angular speed of the propeller is $\omega^*=\omega D/U_\infty = 2\pi n D/U_\infty = 2\pi/J$, and the rotation period is $t U_\infty/D=J$. The simulations are performed in two ways: to collect time-averaged statistics, the outer subdomain is fixed, only the inner sliding region rotates at $\omega$, and a velocity boundary condition is applied on the fairwater surface; to collect phase-averaged statistics, the whole domain rotates at angular speed $\omega$, and velocity boundary condition is applied to the shaft to make it stationary.

A four-stage third-order SSP-RK scheme \cite{spiteri-2002} with a nondimensional time step size of $\Delta t U_\infty/D = 2.5\times 10^{-5}$ is adopted for the time marching. The propeller therefore rotates about $0.01$ degree per time step at the design condition. For spatial discretization, it is known that high-order simulation of turbulent flow may experience instabilities due to aliasing errors \cite{jameson-2012}. We observed such instabilities on the fifth- and above orders. To overcome this issue, we have employed the filter reported in \cite{fischer-2001} (with strength $\alpha=0.05$) to stabilize the simulations. Meanwhile, we compared the design-condition results from the fourth-, the fifth-, and the sixth-order schemes to ensure sufficient resolution. It was observed that the mean loads from the fourth- and the fifth-order schemes have small differences (around $3\%$), but the fifth-order scheme resolves the flow structures with more details. On the other hand, both the mean loads and the flow fields from the fifth- and the sixth-order schemes are almost indistinguishable, which indicates that the fifth-order is the optimum choice considering both accuracy and cost. For this reason, the fifth-order scheme has been used for the simulations in what follows. It is also worth mentioning that all simulations were run for a nondimensional time of $tU_\infty/D=100$, and phase- and time-averaging were performed on the last $85$ time units, which represents approximately 102 revolutions at the design condition.

\section{Results and discussion}
\label{sec:results}

\subsection{Propeller loads}
\label{sec:loads}
The loads on a propeller are measured by the thrust and torque coefficients defined as below,
\begin{equation}
K_T = \frac{T}{\rho n^2 D^4} \qquad\text{and}\qquad K_Q = \frac{Q}{\rho n^2 D^5},
\end{equation}
where $T$ and $Q$, respectively, represent the force and the torque exerted by the fluid on a propeller in the axial direction. The efficiency of a propeller is defined as
\begin{equation}
\eta = \frac{TU_\infty}{2\pi n Q} = \frac{J}{2\pi} \cdot \frac{K_T}{K_Q},
\end{equation}
where the variables have the same meanings as previously explained.

\begin{figure}[H]
\centering
\includegraphics[width=4.5in]{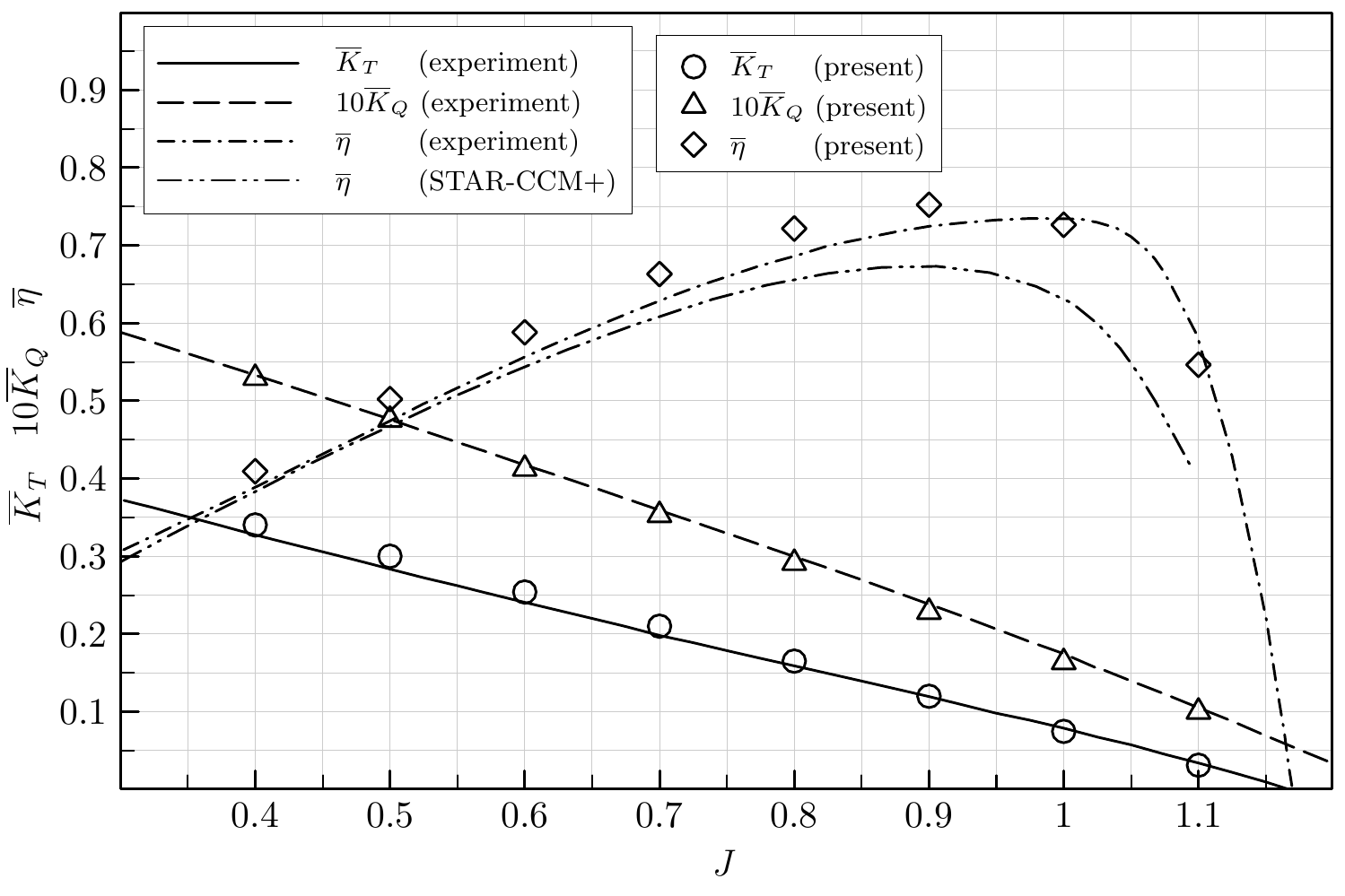}
\caption{The mean blade loads of DTMB 4119 at different working conditions.}
\label{fig:loads}
\end{figure}

We first compare the loads predicted by the simulation with those measured in a previous experiment \cite{jessup-1989} at various advance ratios. As can be seen from Fig. \ref{fig:loads}, the present numerical results agree very well with the experimental values under all working conditions: overloading condition (when $J$ is small), near-design condition, and underloading condition (when $J$ is large). The maximum difference is observed on the efficiency curve, which is around $5\%$. The present simulation predicts the highest efficiency around $J=0.9$, which is close to the design condition that is $J=0.833$ (see Tab. \ref{tab:propellerparam}). The experiment, however, shows an optimum performance slightly above $J=1.0$, which is further away from the design condition. Meanwhile, in the same figure, we also compare the present results with the latest (also the best available) results on the same propeller from a low-order simulation \cite{sezen-2019} using the commercial software STAR-CCM+. It is evident that the low-order solver predicts the efficiency well only in a narrow range of working conditions where $J$ is small. When $J$ increases, the low-order prediction becomes much worse. As will be shown in the next section that when $J$ increases, the flow vortices become weaker, which are more vulnerable to numerical dissipations. Since the high-order method introduces much smaller numerical dissipations, it predicts the loads very accurately under all conditions. In contrast, the large numerical dissipations of the low-order method have completely demolished the predictions when $J$ is large.

\begin{figure}[H]
\centering
\includegraphics[width=5.8in]{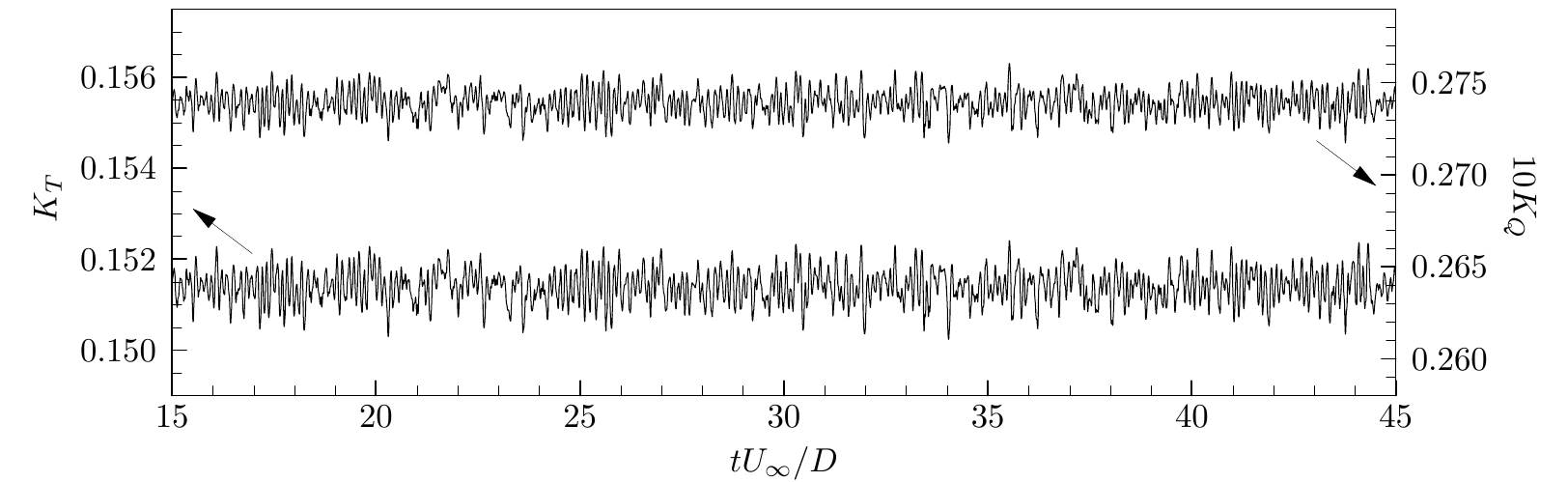}
\caption{Instantaneous loads on the blades of DTMB 4119 at design condition.}
\label{fig:ktkq}
\end{figure}

Since a propeller works around the design condition most of the time, therefore a detailed look into this condition is presented in what follows. The instantaneous $K_T$ and $K_Q$ of the blades at this condition are plotted in Fig. \ref{fig:ktkq} from $tU_\infty/D=15$ to $45$ for about $36$ revolutions. The two coefficients are seen to fluctuate at small amplitudes about their means chaotically due to the turbulent nature of the flow. The mean values (averaged for about 102 revolutions) are compared with the design \cite{denny-1968} and the experimental \cite{jessup-1984,jessup-1989} values in Tab. \ref{tab:ktkq_mean}, where the difference is defined as (simulation/experiment$-1$)$\times 100\%$. The design was based on potential flow theories. The two experiments were performed in open water at slightly different Reynolds numbers. The present Reynolds number is chosen to exactly match that of \cite{jessup-1984},  but it is obvious that the Reynolds number effects are small at this level. Overall, we see very good agreements between the simulation and the experiments as well as the design.
\begin{table}[H]
\setlength{\tabcolsep}{2mm}
\centering
\begin{tabular}{>{\centering}m{1.8cm} >{\centering}m{1.9cm} >{\centering}m{1.8cm} >{\raggedleft}m{2.7cm} >{\raggedleft}m{2.7cm} >{\centering\arraybackslash}m{2.0cm}}
\hline
~ & $Re_D~(\times 10^5)$ & $Re_c~(\times 10^5)$  & $\overline{K}_T$ ~~~(diff.) &  $\overline{K}_Q$ ~~~(diff.) & $\overline{\eta}$ \\
\hline
present                      & 5.59 & 7.3  & 0.1514 \hphantom{(-$2.3\%$)} & 0.0274 \hphantom{(-$2.3\%$)} & 0.7326 \\
design    \cite{denny-1968}  & -    & -    & 0.1540 (-$1.7\%$) & 0.0290 (-$5.5\%$) & 0.7040 \\
~\! exp.~ \cite{jessup-1984} & 5.59 & 7.3  & 0.1500 (~$0.9\%$) & 0.0285 (-$3.9\%$) & 0.6978 \\
~\! exp.~ \cite{jessup-1989} & 7.66 & 10.0 & 0.1460 (~$3.7\%$) & 0.0280 (-$2.1\%$) & 0.6913 \\
\hline
\end{tabular}
\caption{Mean loads on the blades of DTMB 4119 at design condition.}
\label{tab:ktkq_mean}
\end{table}

The simulation allows us to study the loads on different parts of the propeller. We have summarized the results in Tab. \ref{tab:ktkq_mean2}, where ``mean'' denotes the time-averaged values, ``r.m.s'' denotes the root-mean-square of the unsteady components, and the subscripts ``$p$'' and ``$v$'' denote  pressure and viscous contribution, respectively. Before proceeding further, it is worth noting that the thrust on the fairwater needs to be treated carefully. Unlike the blades which are two-sided and closed, the fairwater is ``one-sided'', i.e., it has no direct upstream counterpart to balance the pressure force on its outer surface. In real applications, this force will be partially compensated by the force on an upstream surface with the same cross-sectional area (for example, part of the hull), resulting in a much smaller net force. The flow in the upstream region is usually at reduced speed (or even at stagnation) with higher pressure than that of the freestream. Thus, an underestimated pressure force on this upstream surface is $p_\infty(\pi D_h^2/4)$, where $D_h$ is the hub diameter. We then subtract this force from that on the fairwater to approximate the net thrust on the fairwater.

From Tab. \ref{tab:ktkq_mean2}, we see that only the blades experience a thrust (positive $K_T$), while the hub and the fairwater experience drags (negative $K_T$). On the other hand, all three parts experience positive torques. For the blades, pressure contribution dominates the loads; the {r.m.s.} values are at least three magnitudes smaller than the means, suggesting quasi-steady loads. The $K_T$ and $K_Q$ of the hub are four magnitudes smaller than those of the blades, which indicates that the hub has negligible contribution/effect to the overall performance of the propeller. Furthermore, viscous contribution dominates the hub loads, which is consistent with the fact that the hub has no projection in the axial direction and thus pressure has no way to contribute. In fact, pressure should ideally have zero contribution to the hub loads, and the present very small pressure contribution implies very small geometric imperfection of the hub, which obviously has benefited from the high-order curved representation of the geometry. For the fairwater, the drag is about $2.1\%$ of the thrust on the blades, and pressure dominates; the torque is negligibly small, and viscosity dominates due to the geometric symmetry. Overall, we conclude that the thrust and the torque of the blades, and the drag of the fairwater are the main factors that affect the propeller's performance.
\begin{table}[H]
\setlength{\tabcolsep}{2mm}
\centering
\begin{tabular}{>{\centering}m{1.4cm} m{1.0cm} >{\centering}m{1.3cm} >{\centering}m{1.3cm} >{\centering}m{1.3cm} >{\centering}m{1.3cm} >{\centering}m{1.3cm} >{\centering\arraybackslash}m{1.3cm}}
\hline
                           &        & $K_T$ & $K_{T,p}$ & $K_{T,v}$ & $K_Q$ & $K_{Q,p}$ & $K_{Q,v}$ \\
\hline
\multirow{2}{*}{blades}    & mean   & ~0.1514 & ~0.1519 & -4.9E-4 & 0.0274 & 0.0272\hphantom{2} & 2.4E-4 \\
                           & r.m.s. & ~3.6E-4 & ~3.6E-4 & ~5.6E-7 & 7.0E-5 & 7.0E-5\hphantom{2} & 2.6E-8 \\
\hline
\multirow{2}{*}{hub}       & mean   & -8.5E-5 & -6.7E-7 & -8.5E-5 & 3.9E-6 & 3.1E-8\hphantom{2} & 3.9E-6 \\
                           & r.m.s. & ~2.1E-7 & ~4.5E-8 & ~2.0E-7 & 2.2E-8 & 5.1E-9\hphantom{2} & 2.2E-8 \\
\hline
\multirow{2}{*}{fairwater} & mean   & -3.2E-3 & -3.2E-3 & -2.4E-5 & 8.8E-7 & 6.9E-10            & 8.8E-7 \\
                           & r.m.s. & ~1.2E-4 & ~1.2E-4 & ~4.8E-7 & 9.9E-8 & 9.3E-8\hphantom{2} & 3.2E-8 \\
\hline
\end{tabular}
\caption{Loads on different parts of DTMB 4119 with an ellipsoidal fairwater at design condition.}
\label{tab:ktkq_mean2}
\end{table}

\begin{figure}[H]
\centering
\includegraphics[width=5.4in]{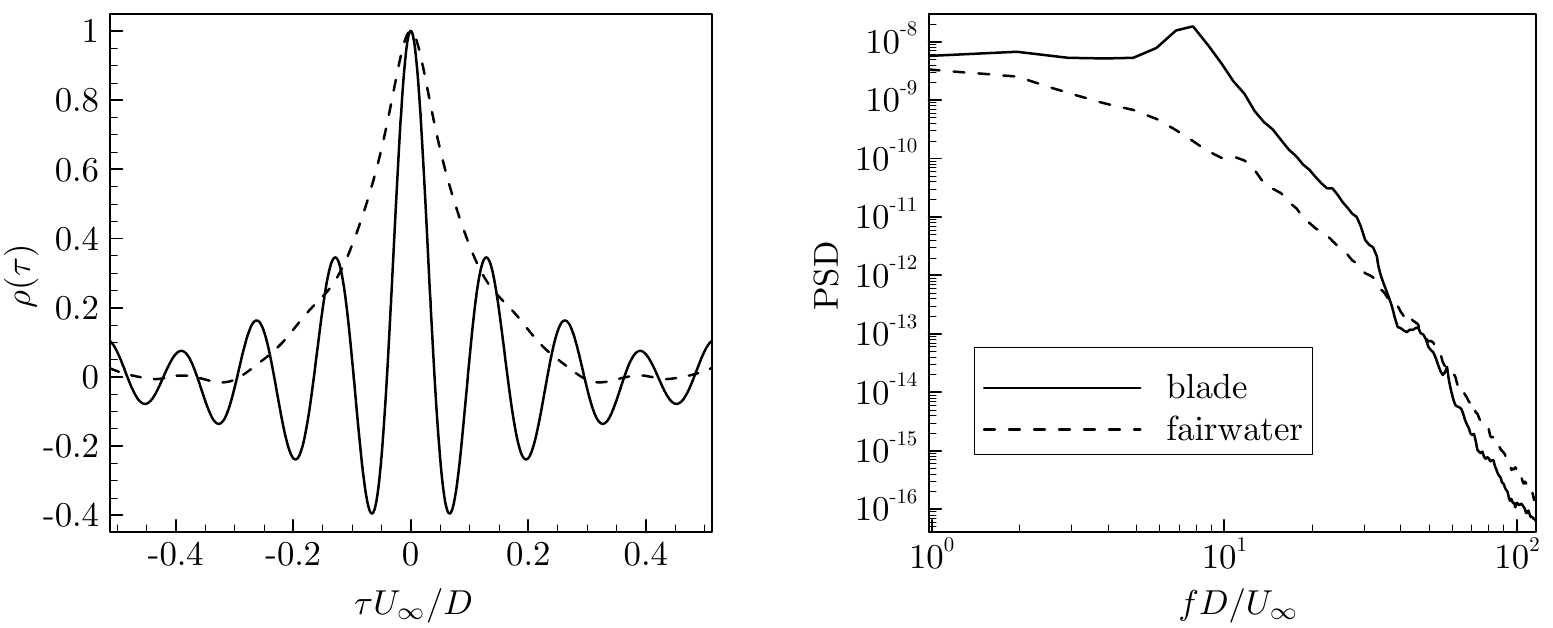}
\caption{Autocorrelation and PSD of the blade thrust and fairwater drag.}
\label{fig:spectra}
\end{figure}

The time and frequency scales of the major loads can be identified through the autocorrelation and the power spectral density (PSD) curves in Fig. \ref{fig:spectra}. It is worth mentioning that the torque and the thrust of a blade have almost identical characteristics, the curves for the torque are therefore not repeated here. The autocorrelation is defined as $\rho(\tau)=R(\tau)/R(0)$, where $R(\tau)=\left<T(t)T(t+\tau)\right>$ is the autocovariance with a time lag $\tau$, and ``$\left<~\right>$'' denotes ensemble average. We calculate the PSD using Welch's method with a $50\%$ overlapping Hanning window and then average the results over 102 revolutions.

The narrow main lobe of the autocorrelation of the blade in Fig. \ref{fig:spectra} implies small integral time scales of the unsteadiness around the blade. It also agrees with the corresponding high-frequency peak around $fD/U_\infty \approx 7.9$ on the PSD curve. In contrast, the autocorrelation of the fairwater has a much wider main lobe, which indicates much larger time scales of the dominant unsteadiness around the fairwater. The corresponding PSD curve is rather broadband and is dominated by very low-frequency components. A comparison of the two PSD curves reveals that the fairwater experiences more unsteadiness at very high-frequencies (e.g., $fD/U_\infty > 20$) than the blade does. These time and frequency scales are directly related to flow structures which will be discussed in the next section.

\subsection{Flow fields}
The flow fields of a propeller have very distinct flow structures that are of crucial importance to the propeller's performance. This has made studying the formation, mutual interaction, and stability of these flow structures a constant research topic for decades. In this section, we report the details of the flow fields of DTMB 4119, including the vortices, the velocity field, and the pressure field.

\subsubsection{Vortices}
\label{sec:vortices}

The vortical structures in a flow field can be well visualized by isosurfaces of Q-criterion \cite{hunt-1988}. The Q-criterion (denoted by $Q_{cr}$) is defined as the second invariant of the velocity gradient tensor, i.e., $Q_{cr}=(\Omega_{ij}\Omega_{ij} - S_{ij}S_{ij})/2$, where $\Omega_{ij}=(u_{i,j}-u_{j,i})/2$ and $S_{ij}=(u_{i,j}+u_{j,i})/2$ are the antisymmetric and the symmetric component, respectively, of the velocity gradient tensor.

\begin{figure}[H]
\centering
\includegraphics[width=\textwidth]{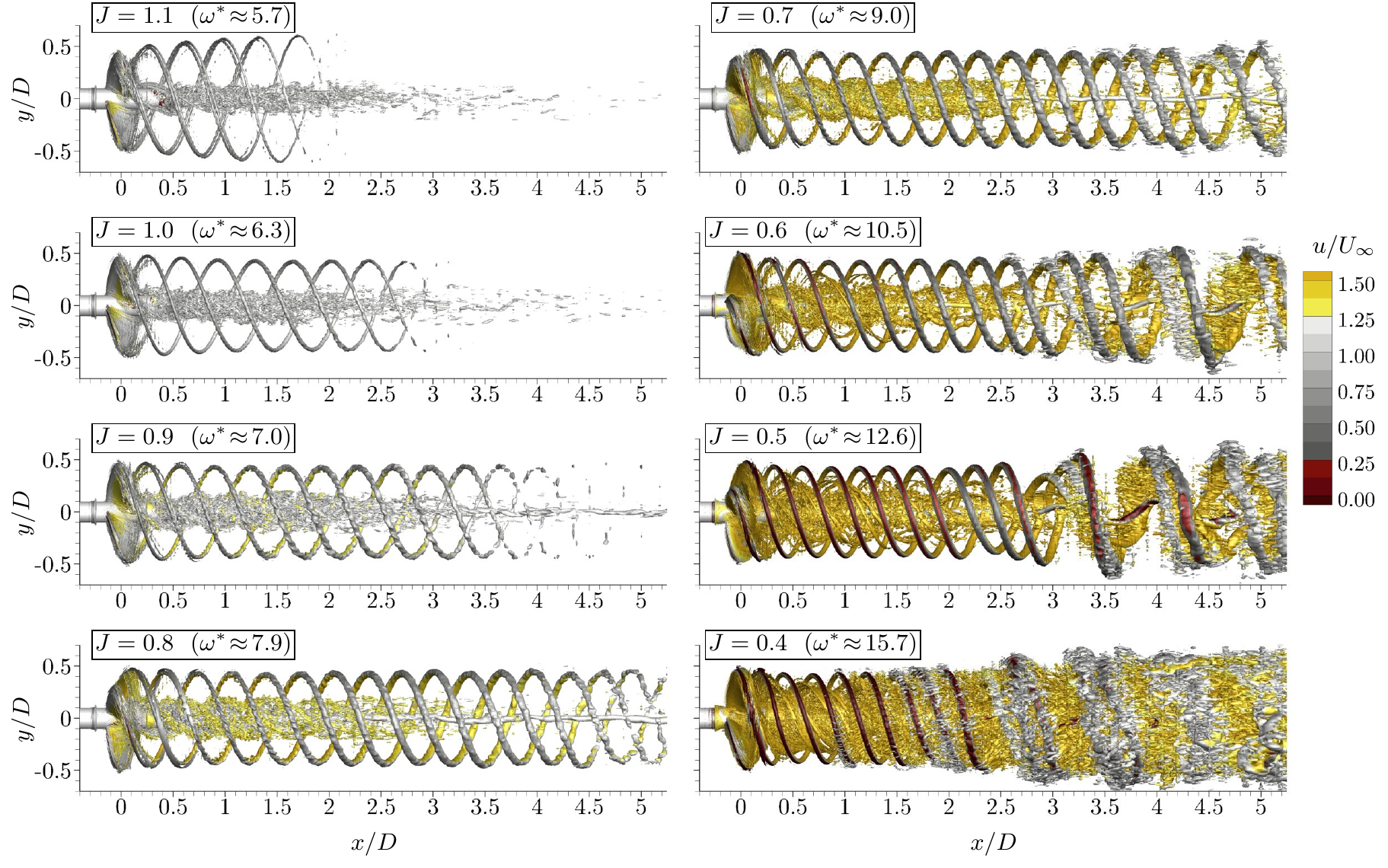}
\caption{Isosurfaces of $Q_{cr}D^2/U_\infty^2=40$ at different working conditions.}
\label{fig:ellip_instant_Qcrs}
\end{figure}

When the Reynolds number is given, the only parameter that determines a propeller's flow field is the advance ratio $J$. Figure \ref{fig:ellip_instant_Qcrs} shows the instantaneous vortical structures as $J$ decreases from $1.1$ to $0.4$. Note that the nondimensional rotational speed is related to the advance ratio as $\omega^*\!= 2\pi/J$. Thus, a decreasing $J$ is equivalent to an increasing $\omega^*$.

We notice two dramatic changes in the flow field as $J$ decreases: the increase of vortex strength and the occurrence of flow instabilities. At $J=1.1$ and $1.0$, the vortices are so weak that they are quickly dissipated by the wake flow. At $J=0.9$ and $0.8$, the vortices become strong enough to sustain for a long distance in the wake, and a hub vortex is also well established. In addition, up to this point the flow remains stable. Obvious instability occurs when $J$ decreases to $0.7$, and the instability is caused by mutual interactions between two tip vortices around $x/D=4.8$. At $J=0.6$, the instability is still caused by mutual tip vortex interactions, but the occurrence moves upstream to $x/D=3.4$. The occurrence further moves upstream to $x/D=3.2$ and $2.7$, for $J=0.5$ and $0.4$, respectively. However, the cause of the instability becomes more complicated. At $J=0.5$, it seems the instability not only comes from the mutual interaction between the tip vortices, but also the interaction between tip and hub vortices. Finally, at $J=0.4$, it looks like the trailing edge vortices have become strong enough to be the leading cause of the instability. It was conjectured in \cite{kumar-2017} that blade trailing edge vortices are an important source of flow instabilities. Based on our observations here, this is only possible when $J$ is small enough (i.e., propeller is at very high relative rotational speed) and when blade trailing edge vortices are strong enough.

\begin{figure}[H]
\centering
\includegraphics[width=5.2in]{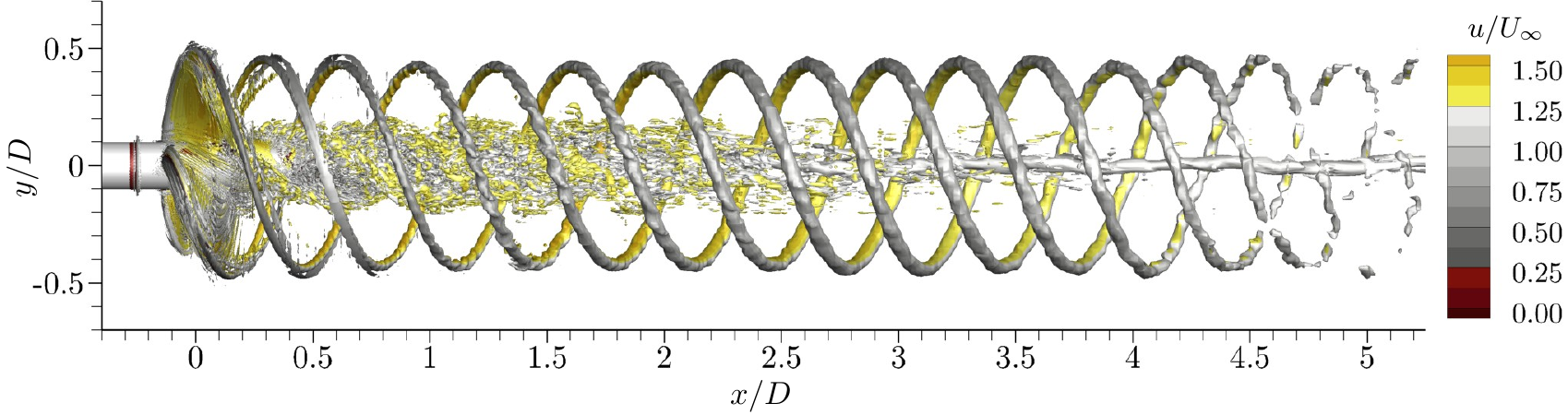} \\
\includegraphics[width=5.2in]{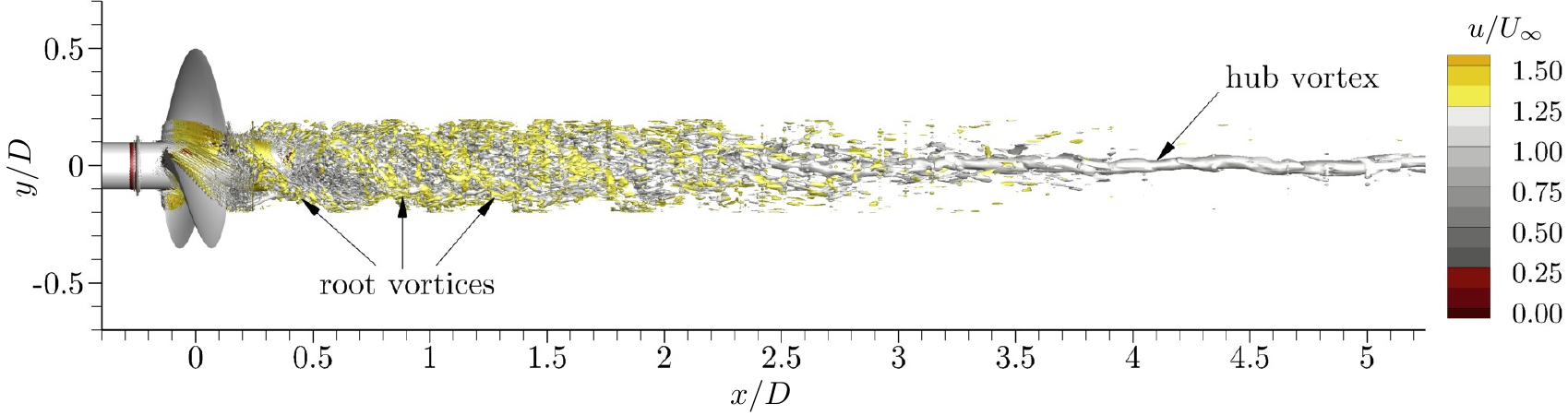}
\caption{Isosurface of instantaneous Q-criterion $Q_{cr}D^2/U_\infty^2=40$ at design condition.}
\label{fig:ellip_instant_Qcr}
\end{figure}

We already saw that the flow fields can be very different at different working conditions. In the rest of this paper, we focus on the design condition only. Figure \ref{fig:ellip_instant_Qcr} shows an instantaneous view of the flow structures at this condition. It is seen that the tip vortices are very much equally spaced along the axial direction, with the distance between two successive vortices being approximately $0.36D$, which is about one-third the tip pitch (see Tab. \ref{tab:propellergeo}). Meanwhile, the surface velocity contours reveal that a tip vortex has lower streamwise speed on the outer surface, and higher speed on the inner surface. This indicates that a tip vortex not only revolves helically about the propeller's axes, but also about its own core at the same time. The topology of the root vortices are not very obvious from this instantaneous flow field due to the many small turbulent structures. These high-frequency small flow structures are closer to the fairwater than to the blades, and thus have contributed more to the load unsteadiness of the fairwater than to the blades, which agrees with our previous observation on the PSD curves in Fig. \ref{fig:spectra}. These small structures, however, do not dominate the load unsteadiness, which is likely because of their isotropy that leads to mutual cancellation of the effects. While the tip vortices break up about $5D$ downstream from the propeller, the hub vortex stays strong and does not break up even at the outlet (i.e., $12D$ downstream from the propeller; complete picture not shown here due to limited space).

\begin{figure}[H]
\centering
\includegraphics[width=5.2in]{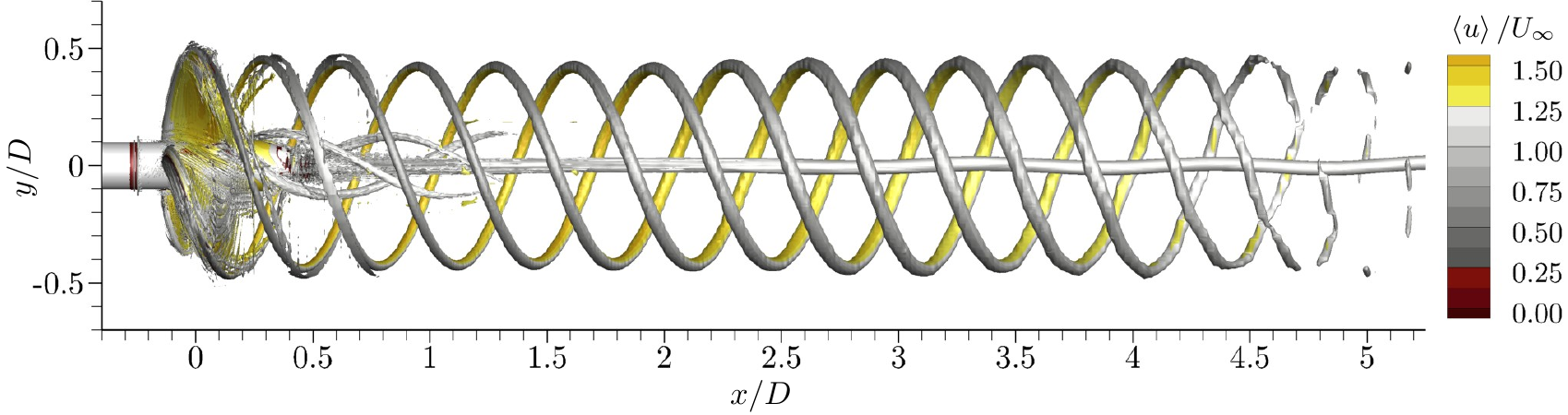} \\
\includegraphics[width=5.2in]{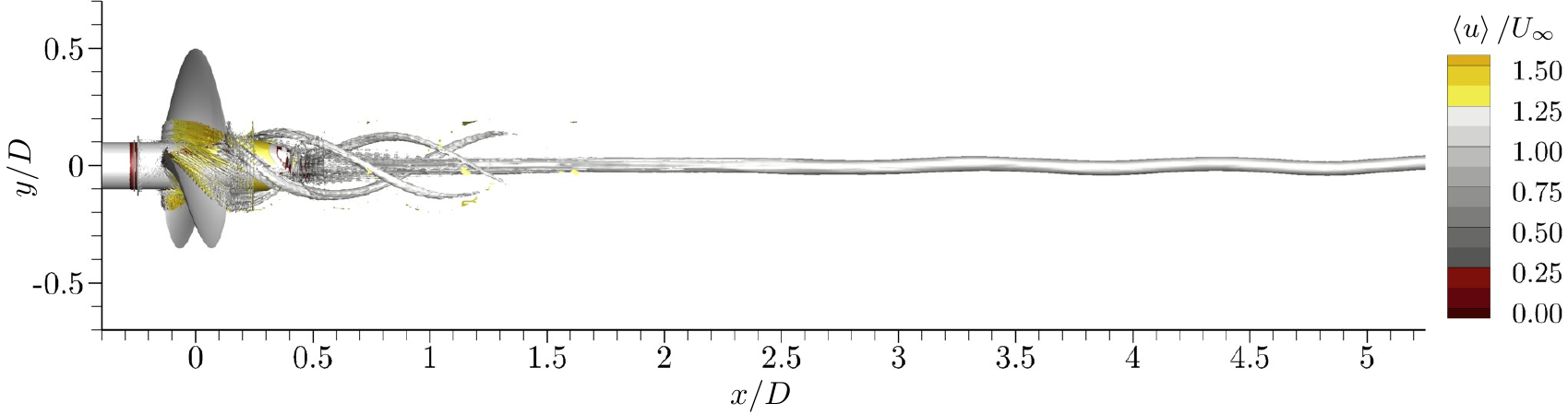}
\caption{Isosurface of phase-averaged Q-criterion $\left<Q_{cr}\right>D^2/U_\infty^2=40$ at design condition.}
\label{fig:ellip_phase_Qcr}
\end{figure}

The phase-averaged isosurfaces of Q-criterion are shown in Fig. \ref{fig:ellip_phase_Qcr}, and they reveal the major flow structures, especially the root vortices, more evidently. The phase-averaged hub vortex is still seen to vary along the axial direction. In fact, it is the variations of these big structures that dominate the load unsteadiness of the fairwater. Similarly, the load unsteadiness of the blades is likely dominated by the unsteadiness of the tip and the trailing edge vortices.

The FR method is a discontinuous-type of method, and reconstruction must be performed to make the solutions and fluxes globally continuous. This rule also applies to the statistics, which are only element-wise continuous unless reconstructed. However, reconstructing the statistics will impose extra computational and memory cost to the simulation. For this reason, this process is not performed in this work, which results in non-smoothness across cell boundaries as can be seen from Fig. \ref{fig:ellip_phase_Qcr}, especially in the vicinity of the propeller where flow changes rapidly. Nevertheless, this simplification should not alter any of the conclusions here.

\begin{figure}[H]
\centering
\includegraphics[width=5.2in]{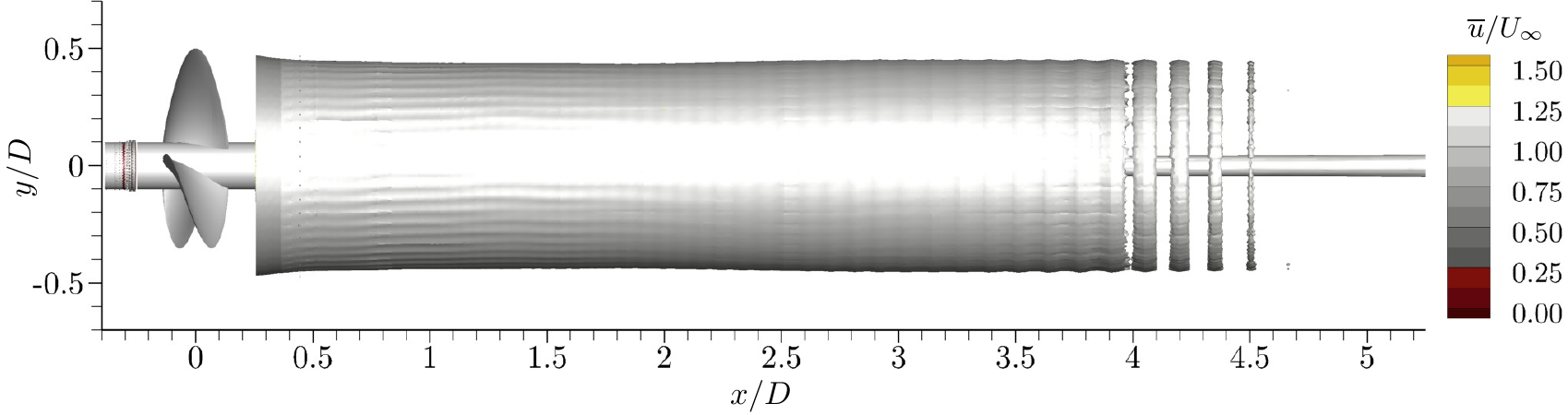} \\
\includegraphics[width=5.2in]{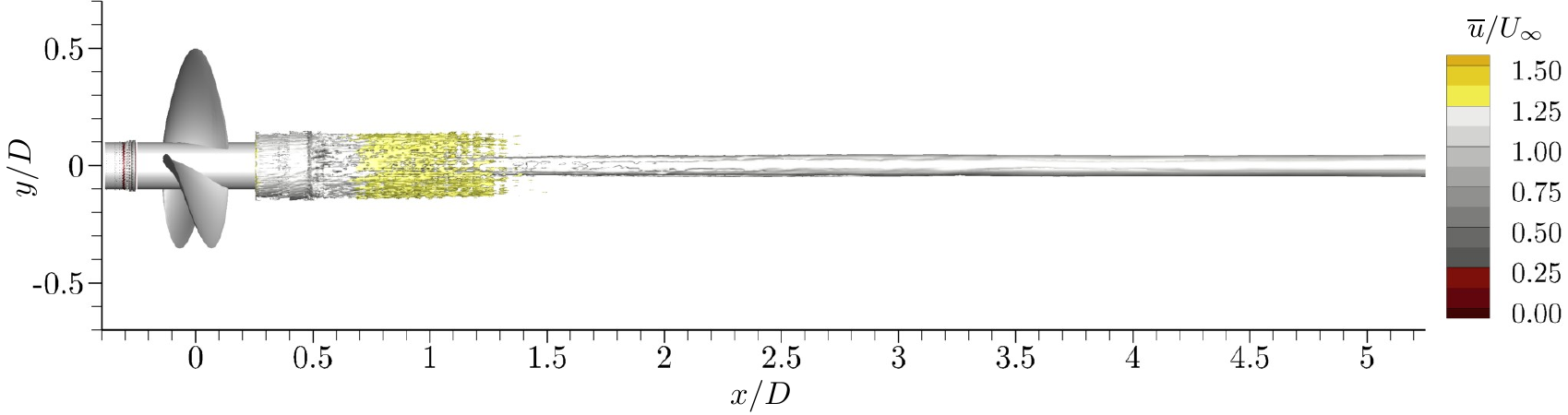} \\
\includegraphics[width=5.2in]{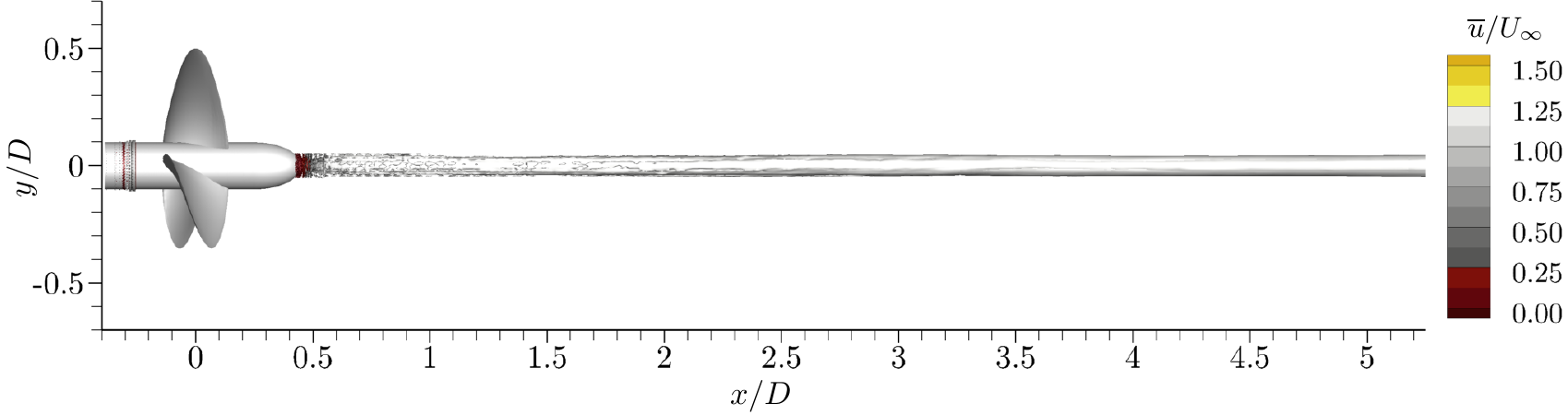}
\caption{Isosurface of time-averaged Q-criterion $\overline{Q}_{cr}D^2/U_\infty^2=8$ at design condition.}
\label{fig:ellip_time_Qcr}
\end{figure}

The time-averaged flow field is shown in Fig. \ref{fig:ellip_time_Qcr}. It is worth noting that time-averaging is impossible for the sliding region due to the movement of the propeller, and the flow in this region is thus not shown in the figure. Over time, the tip vortices form a slightly converging-diverging ``duct'' in space. The root vortices, because of the very small instantaneous turbulent structures, are very difficult to converge in time. Nevertheless, they still have a tube-like shape over time in space. Unlike the instantaneous and the phase averaged ones, the time-averaged the hub vortex is very symmetric about the axis, and almost sees no deviation from the axial direction. This clearly demonstrates that time-averaging not only helps remove most of the small unsteadiness, but also the large ones, from the flow field.

\begin{figure}[H]
\centering
\includegraphics[width=5.2in]{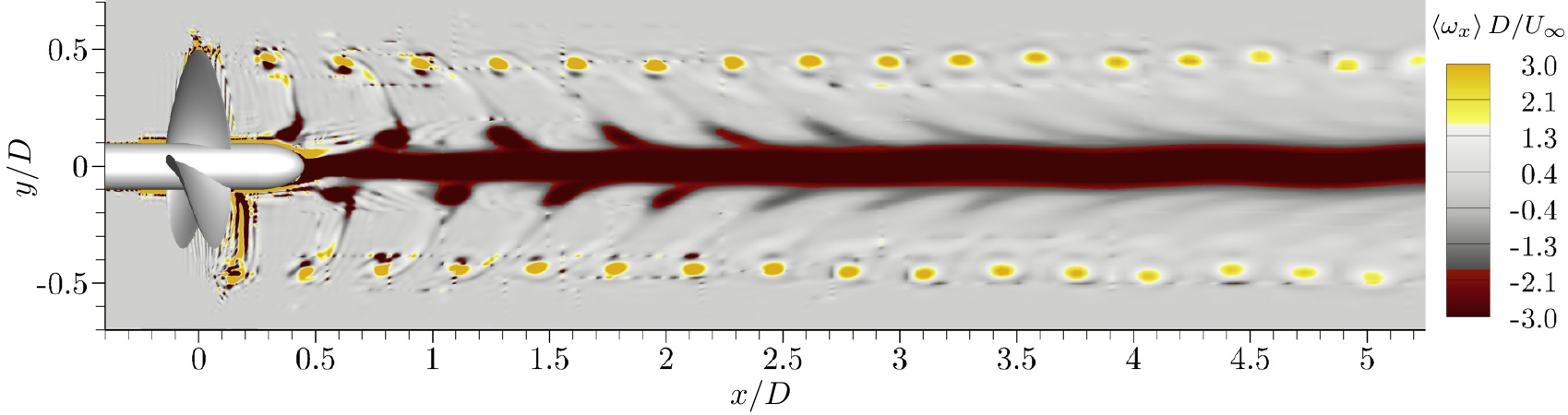}
\caption{Contours of phase-averaged streamwise vorticity in the central $x$-$y$ plane.}
\label{fig:ellip_phase_vort}
\end{figure}

\begin{figure}[H]
\centering
\includegraphics[width=5.8in]{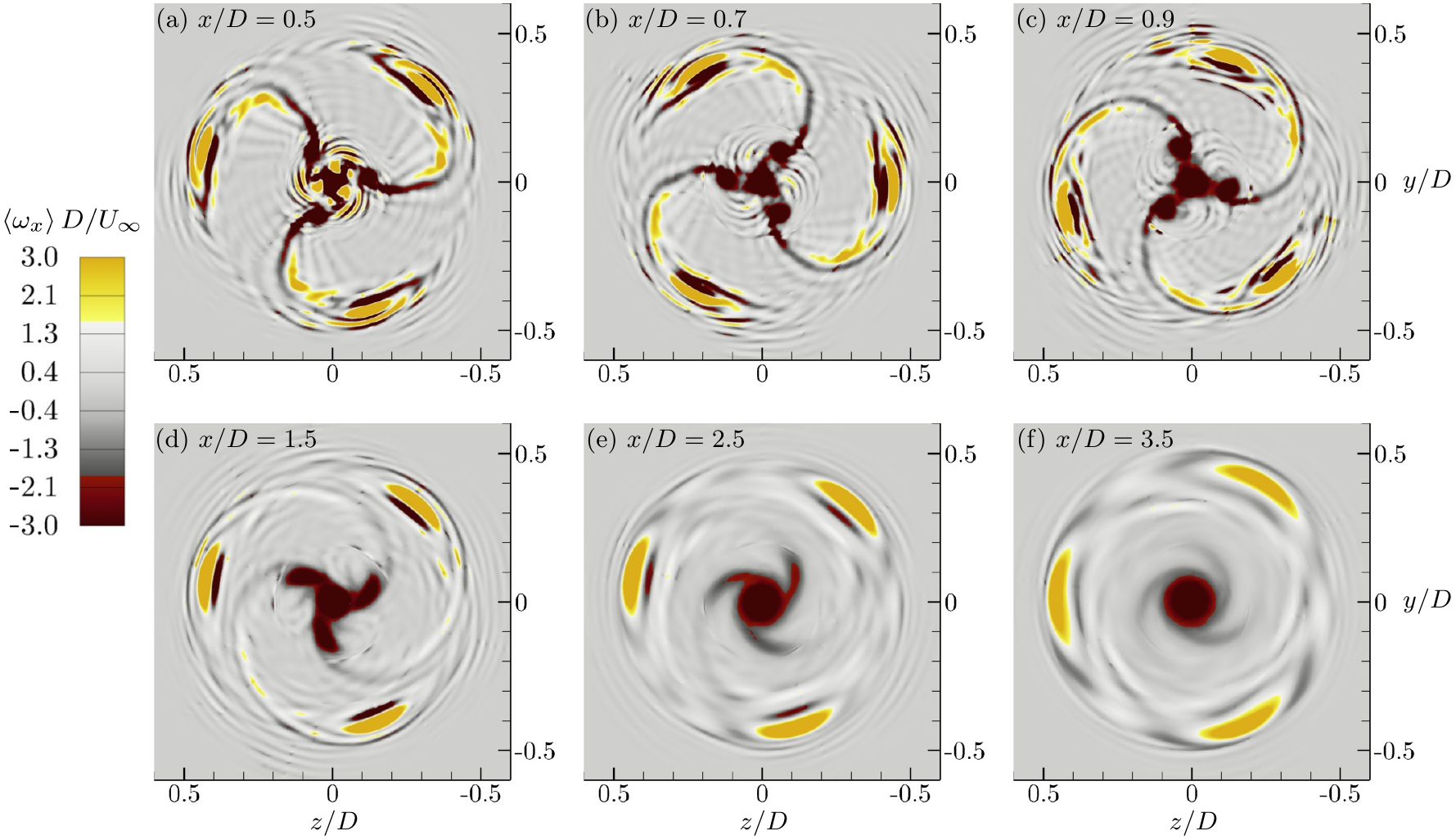}
\caption{Contours of phase-averaged streamwise vorticity in $y$-$z$ planes at different streamwise locations.}
\label{fig:ellip_phase_vort_yz}
\end{figure}

The Q-criterion isosurfaces are able to reveal the most coherent vortical flow structures. They are, however, inefficient to expose the weak ones like the trailing edge vortices at the design condition. Additionally, the Q-criterion cannot reveal the sign of a vortex. For this reason, we have plotted the streamwise vorticity contours in Figs. \ref{fig:ellip_phase_vort} and \ref{fig:ellip_phase_vort_yz} to fill these gaps.

We can clearly see the footprints of the trailing edge vortices (TEVs) in Fig. \ref{fig:ellip_phase_vort}. One end of each TEV connects to a tip vortex, and the other end connects to the hub or root vortex. As the flow goes downstream, the TEVs tilt more and more towards downstream, which obviously complies with the wake velocity distribution. The signs of the vortices reveal that the tip vortices rotate in the opposite direction to all other vortices as well as the propeller.

The vorticity contours in Fig. \ref{fig:ellip_phase_vort_yz} uncover how the vortices develop in the azimuthal and radial directions as the flow travels downstream. The evolution of the TEVs is the most prominent: they are elongated and bent in the azimuthal direction and finally impinge onto the tip vortices. However, because of flow dissipation and the weak strength of the TEVs, these interactions do not destabilize the flow. Again, the change fo the TEVs is closely related to the velocity distribution that will be discussed in the next section.

\subsubsection{Velocity field}
\label{sec:velocity}

\begin{figure}[H]
\centering
\includegraphics[width=5.2in]{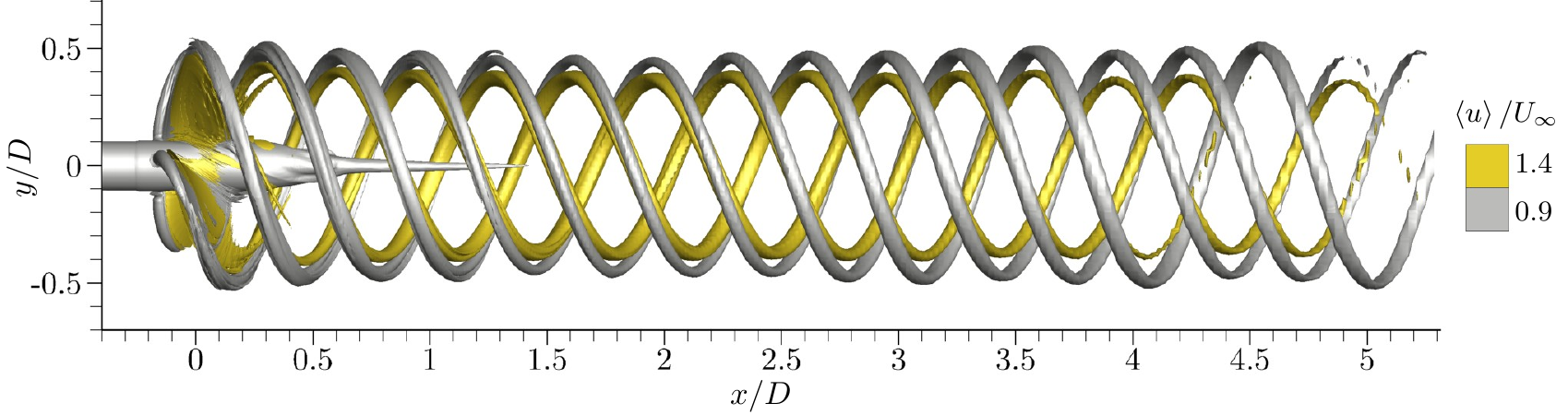}
\caption{Isosurfaces of phase-averaged streamwise velocity $\left<u\right>/U_\infty=0.9$ (gray) and $1.4$ (yellow).}
\label{fig:ellip_uiso1}
\end{figure}

We already noticed that the outer and inner surfaces of a tip vortex have different speeds. The tip vortices thus should be well bounded by velocity isosurfaces. To confirm this, we plot two isosurfaces of the phase-averaged streamwise velocity in Fig. \ref{fig:ellip_uiso1}. It is seen that the isosurfaces of $\left<u\right> /U_\infty = 0.9$ and $1.4$ follow the trajectories of the tip vortices very well. Meanwhile, the increasing gap between the two isosurfaces also agrees with the decreasing strength of the tip vortices as the flow moves downstream.

A closer look of the velocity isosurfaces in the very vicinity of the blades also reveals the formation of the tip vortices. In Fig. 20, we are looking towards downstream at the suction side in (a1) and (a2), and towards upstream at the pressure side in  (b1) and (b2). From  (a1), it is obvious that each leading edge (LE) decelerates the incoming flow, resulting in a strand of low-speed flow along the LE and finally sheds off around the tip. From (a2), each trailing edge (TE) accelerate the flow and sheds off a strand of high-speed flow slightly below the tip. When these two strands of flow meet and be convected downstream, a helical tip-vortex system is generated.
\begin{figure}[H]
\centering
\includegraphics[width=6.2in]{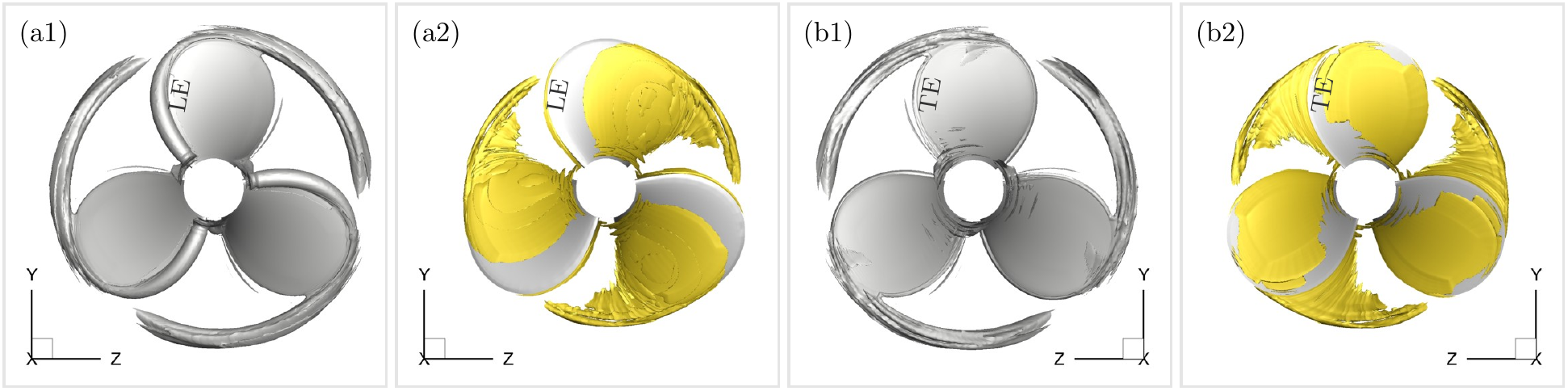}
\caption{Isosurfaces of $\left<u\right>/U_\infty=0.9$ (gray) and $1.4$ (yellow): (a1,a2), suction side; (b1,b2), pressure side.}
\label{fig:ellip_uiso2}
\end{figure}

\begin{figure}[H]
\centering
\includegraphics[width=5.2in]{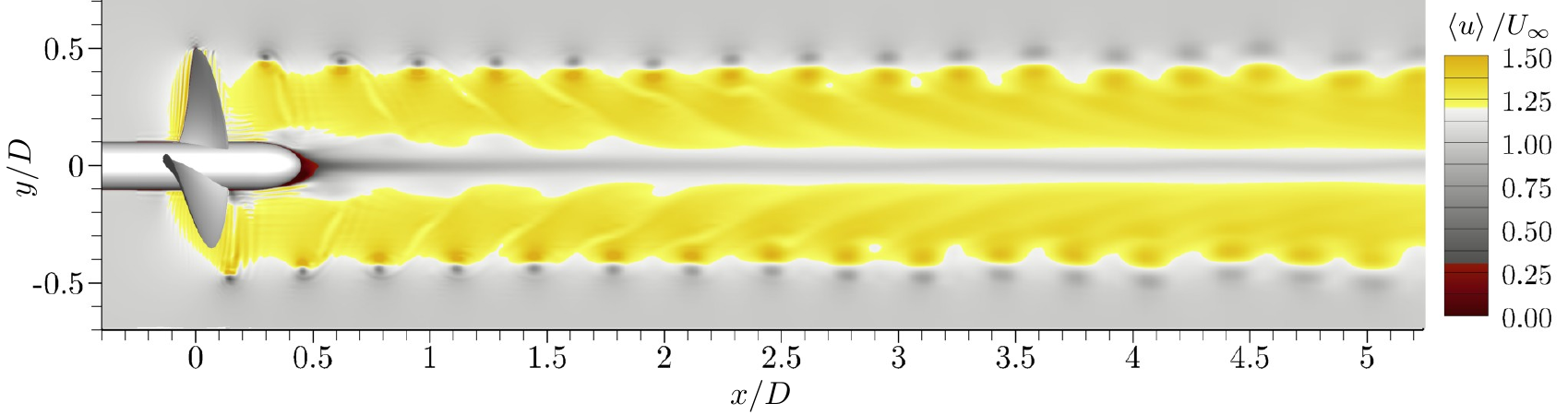}
\caption{Contours of phase-averaged streamwise velocity in the central $x$-$y$ plane.}
\label{fig:ellip_u_xy}
\end{figure}

Figure \ref{fig:ellip_u_xy} shows the phase-averaged streamwise velocity contours in the central $x$-$y$ plane (i.e., $z=0$). It is seen that the blade wake is overall accelerated, while the fairwater wake is mostly at reduced speed. The flow immediately downstream of the fairwater has very low speed, suggesting that a bubble is likely formed in this region. The tip vortices show up as local velocity min-max pairs along the outer edge of the slipstream. The footprints of other vortices, such as the TEVs and the hub vortex, are also visible on the velocity contours.

\begin{figure}[H]
\centering
\includegraphics[width=5.2in]{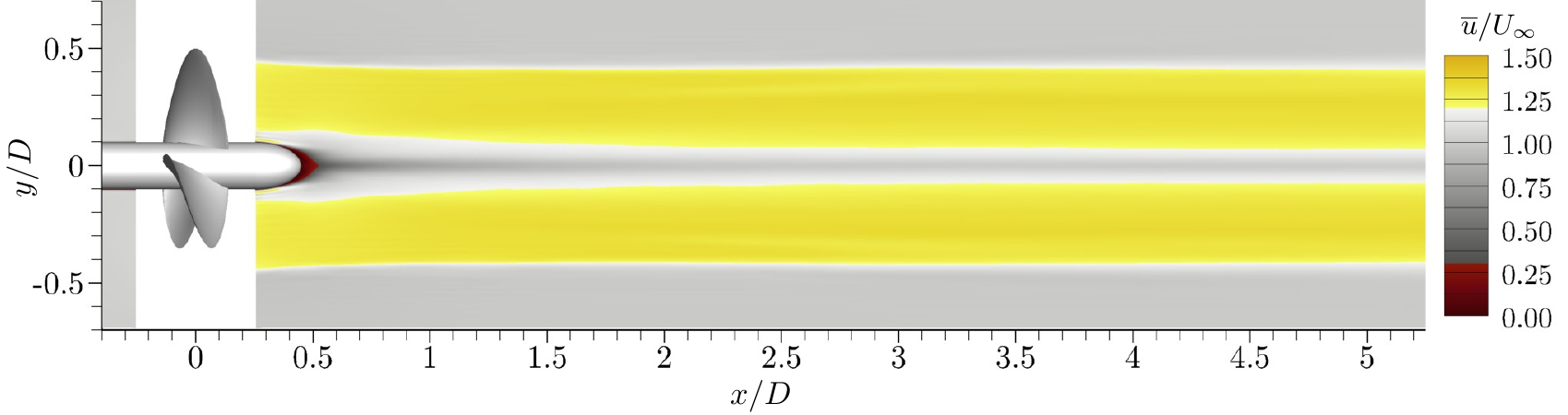} \\
\includegraphics[width=5.2in]{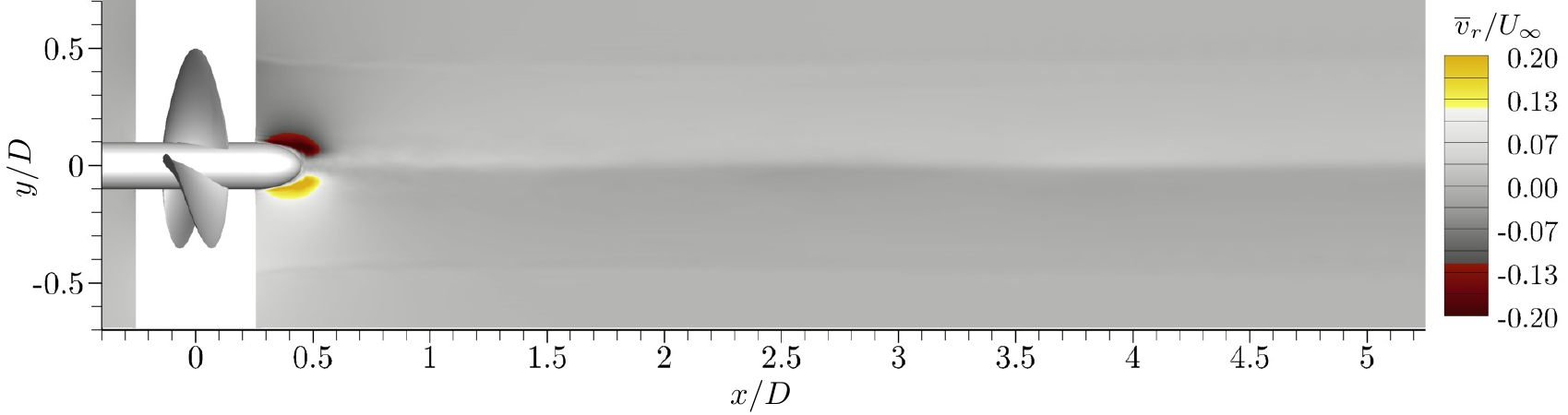} \\
\includegraphics[width=5.2in]{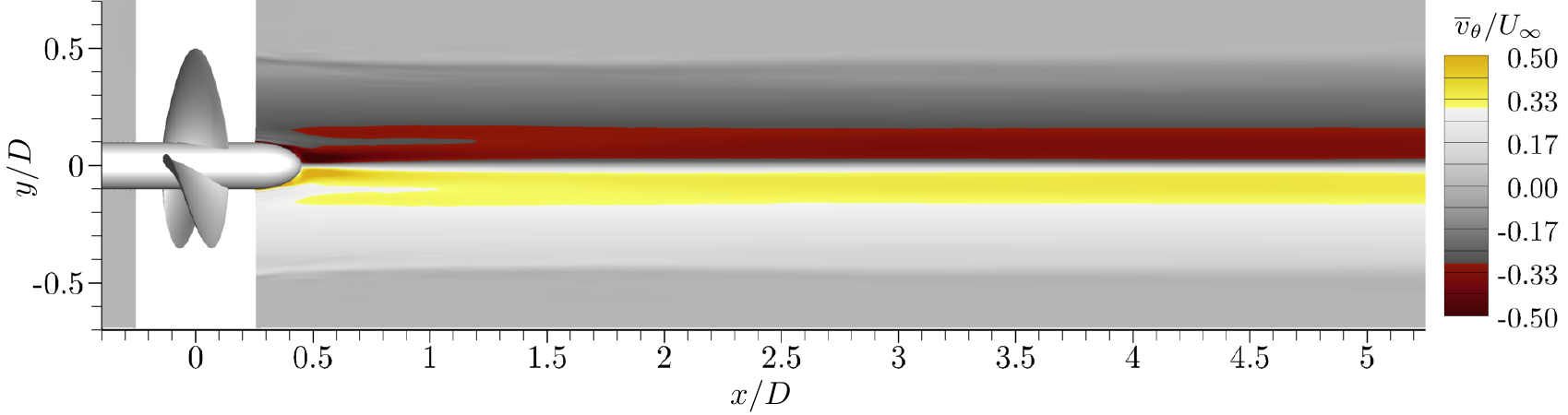}
\caption{Contours of time-averaged streamwise, radial and azimuthal velocity components in the central $x$-$y$ plane.}
\label{fig:ellip_uvw_xy}
\end{figure}

The Cartesian velocity can be decomposed into three components: streamwise, radial, and azimuthal, denoted by $u$, $v_r$, and $v_\theta$, respectively. Figure \ref{fig:ellip_uvw_xy} shows the time-averaged contours of these components. We see that the overall slipstream has a converging-diverging shape and is well contained in the propeller's swept area, i.e., $r \le 0.5D$ (note that $r=|y|$ in the central plane). The radial velocity is small almost everywhere, except in the very vicinity of the fairwater. The azimuthal speed is large only in the fairwater wake and is induced by the strong hub vortex. The two low-speed strips on the azimuthal speed contours around $r/D=0.1$ and $0.4 \le x/D \le 1.2$ are footprints of the root vortices.

\begin{figure}[H]
\centering
\includegraphics[height=2.65in]{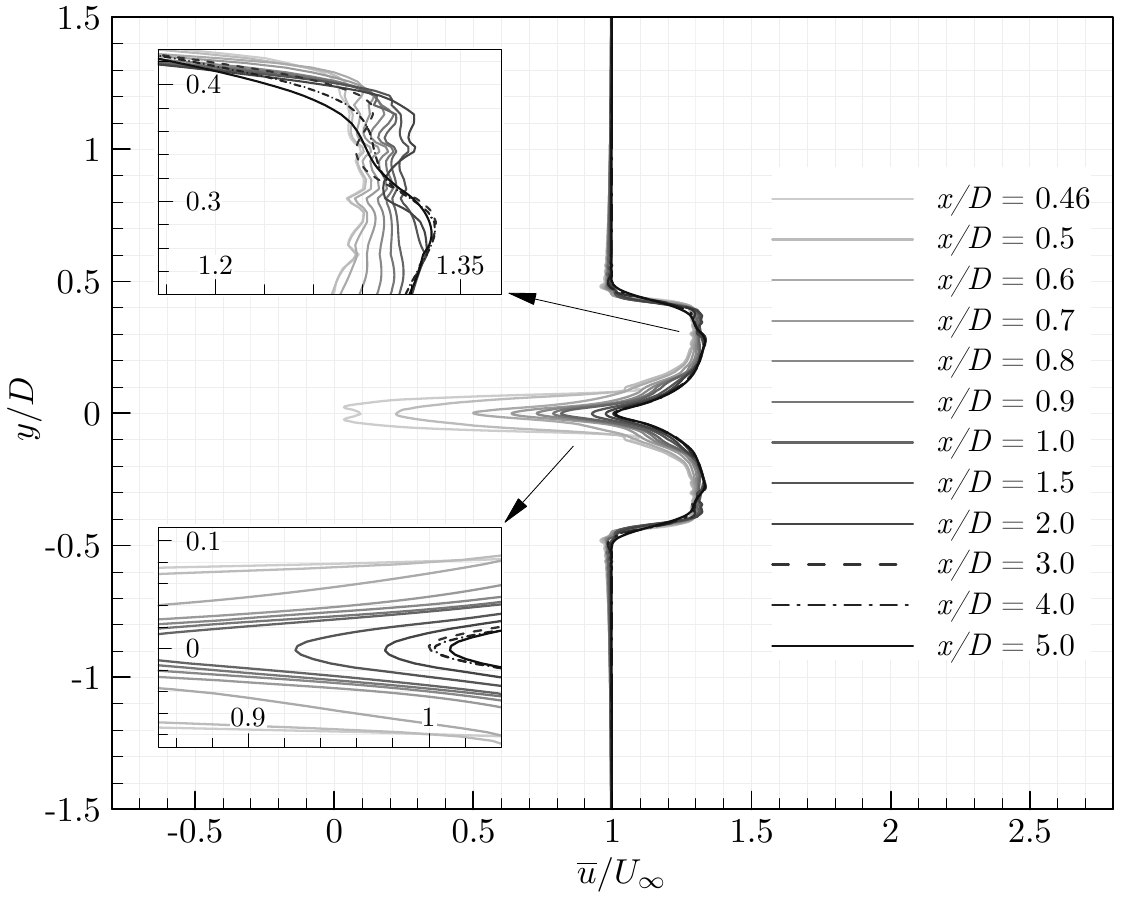}
\includegraphics[height=2.65in]{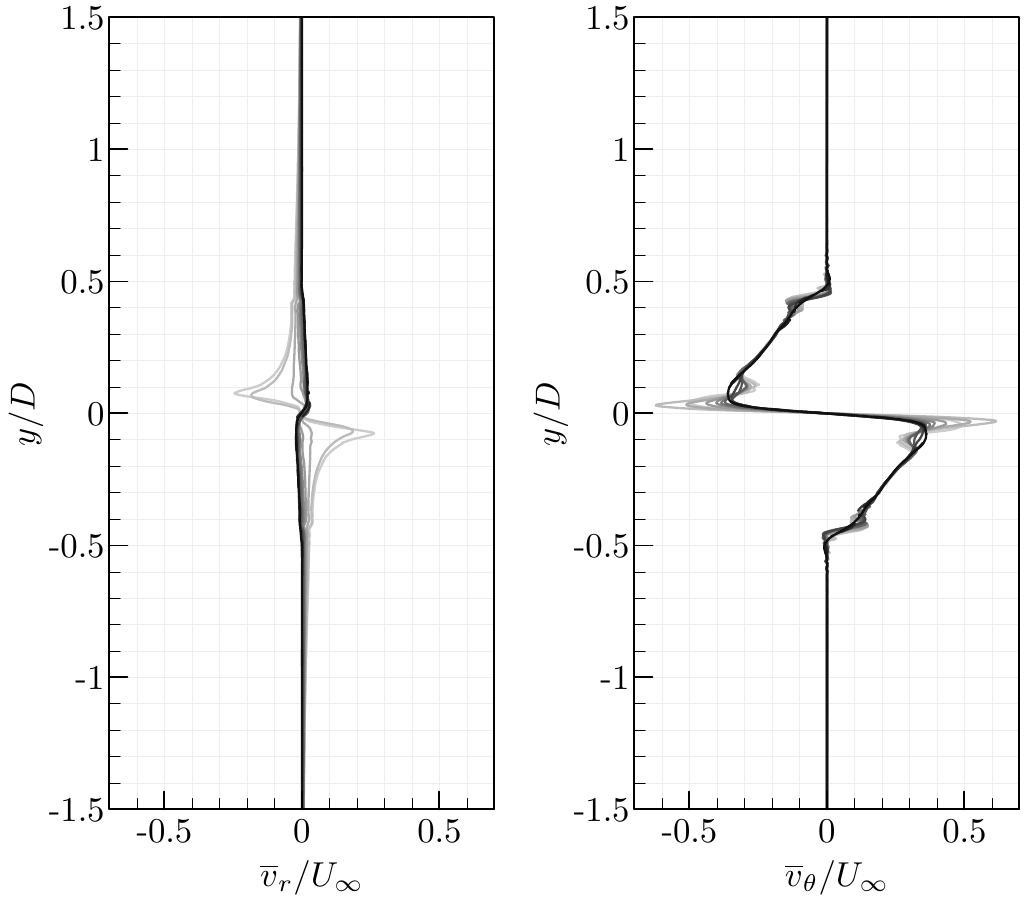}
\caption{Time-averaged velocity profiles at different streamwise locations in the central plane.}
\label{fig:ellip_time_uvw1d}
\end{figure}

More detailed velocity profiles are shown in Fig. \ref{fig:ellip_time_uvw1d}. We see almost no induced velocity outside the slipstream (i.e., $r/D>0.5$) on all profiles, which suggests that the propeller introduces very little disturbance to the flow outside its swept area. The streamwise velocity $\overline{u}$ reaches its maximum values in the region $0.2 < r/D < 0.4$, where the flow is accelerated by more than $30\%$ over $U_\infty$. Outside this region, $\overline{u}$ quickly decreases to $U_\infty$ around $r/D=0.5$, and also decreases to the hub-vortex-core speed at $r/D=0$. This velocity distribution leads to the TEV deformation that is observed in Fig. \ref{fig:ellip_phase_vort}. The streamwise hub-vortex-core speed is close to zero in the near wake region (e.g., $x/D=0.46$), and then consistently increases towards downstream. At $x/D=5.0$, it is already slightly over $U_\infty$. The radial velocity $\overline{v}_r$ is only noticeable in the near wake, for example, at $x/D=0.46$ and $0.5$, and then quickly drops to very small values as the flow travels downstream. The azimuthal velocity $\overline{v}_\theta$ is mostly induced by the hub vortex. It has almost the same profile at different locations, except in two small regions: one around $r/D=0.1$ that is affected by the root vortices, and the other around $r/D=0.42$ that is affected by the tip vortices. The profiles of $\overline{v}_\theta$ are very close to that of a typical Rankine vortex, and is responsible for the TEV deformation in Fig. \ref{fig:ellip_phase_vort_yz}. The maximum value of $\overline{v}_\theta$ is about $0.62U_\infty$ in the near field, and $0.36U_\infty$ in the intermediate wake. These are very large values, and clearly indicate how strong the hub vortex is.

\begin{figure}[H]
\centering
\includegraphics[width=4.2in]{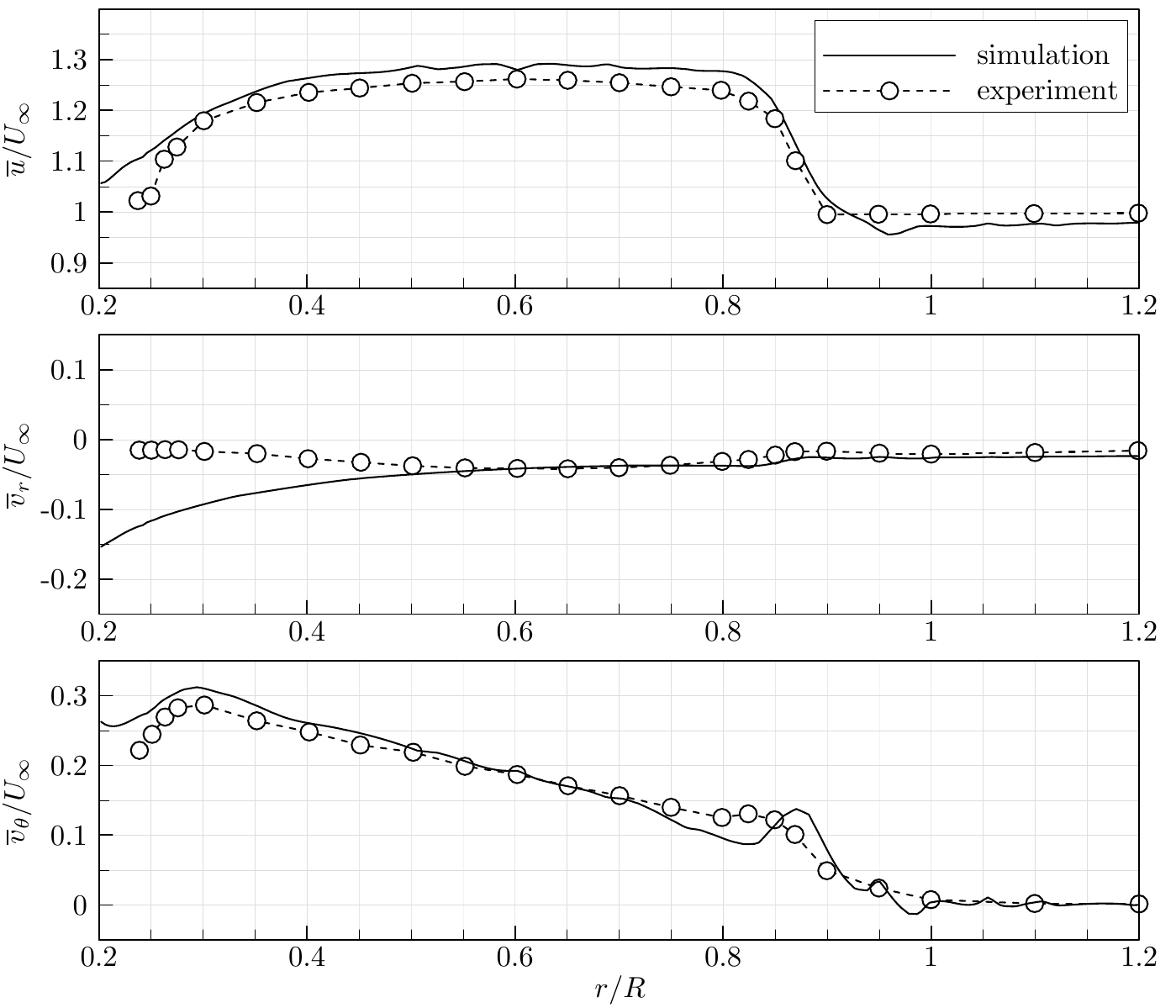}
\caption{Comparison of mean velocity profiles at $x/R=0.951$ with experiment.}
\label{fig:ellip_time_vcompare}
\end{figure}

As a further validation of the simulation, in Fig. \ref{fig:ellip_time_vcompare} we compare the velocity profiles at $x/R=0.951$ with a water tunnel measurement  from \cite{jessup-1989}. Overall, very good agreements are seen between the simulation and the experiment. For example, in the region $0.3 <r/R<1.2$, the maximum difference on $\overline{u}$ is only around $2\%$. However, we do see large discrepancies in the region that is close to $r/R=0.2$. This is because of the setup difference between the experiment and the simulation. In the experiment, the propeller shaft is actually at downstream, making it a stationary wall surface at $(x/R,r/R)=(0.951,0.2)$. In contrast, for the simulation, the shaft is at upstream, and it is a flow region at the same location.

\begin{figure}[H]
\centering
\includegraphics[width=5.8in]{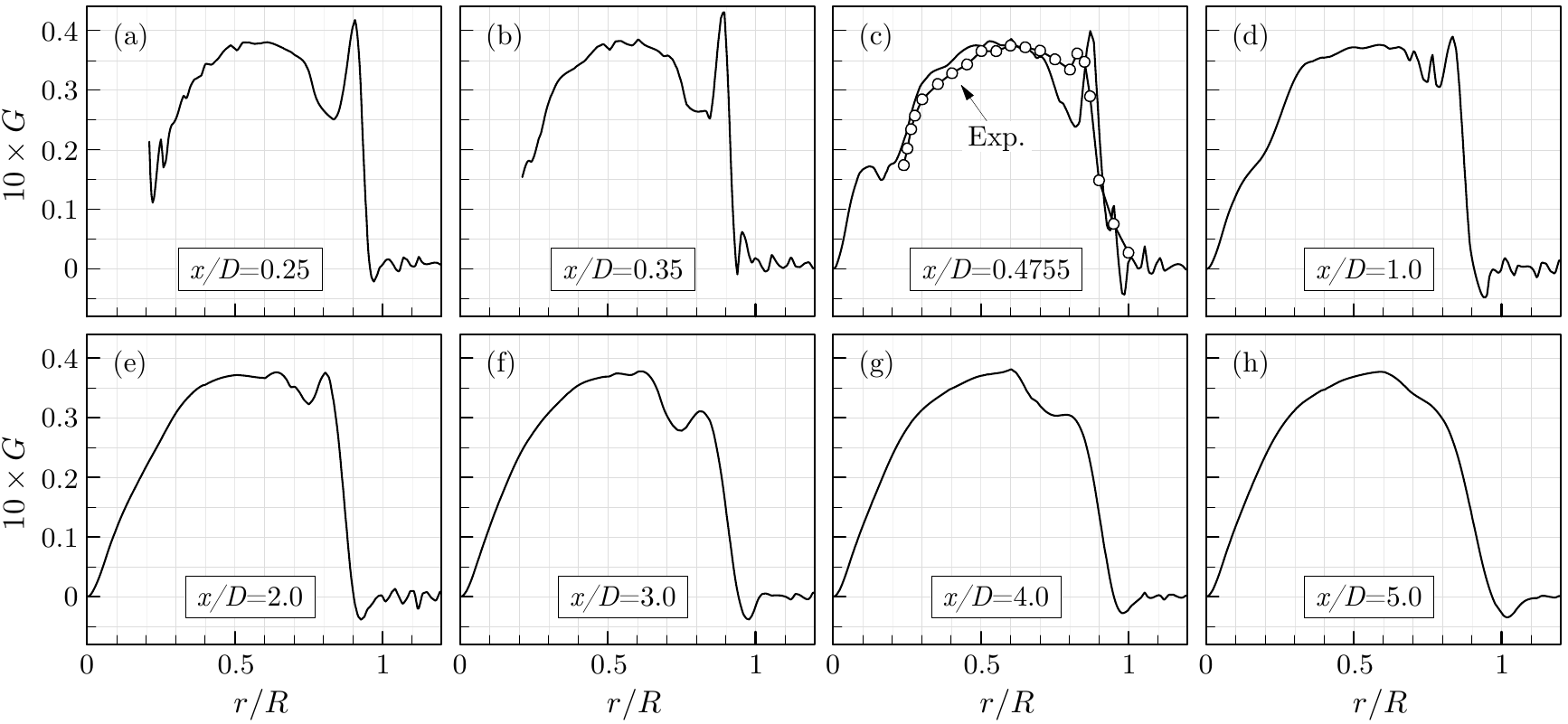}
\caption{Flow circulation at different streamwise locations.}
\label{fig:ellip_time_circu}
\end{figure}

Based on the Kutta-Joukowski theorem, for an inviscid flow the ``lift'' on a unit span of a body (such as a propeller blade) is proportional to the circulation. Following the derivation in \cite{kerwin-1982,wang-1985}, the circulation around a blade section at $r$, denoted by $\Gamma(r)$, is related to the circumferential speed of the slipstream as
\begin{equation}
\Gamma(r) \approx - 2\pi r v_\theta(r)/Z,
\end{equation}
where $Z$ is the number of blades of a propeller. Applying the above relation to the mean flow, we can define the following nondimensional circulation for a blade,
\begin{equation}
G(r) = -\frac{1}{Z} \frac{r}{R} \frac{\overline{v}_\theta(r)}{U_\infty} \approx \frac{\Gamma(r)}{2\pi R U_\infty}.
\end{equation}
This variable can be employed to measure the load distribution on a blade. Moreover, $G$ is conservative for inviscid flows, and should be roughly conservative for high Reynolds number flows (where viscous effects are small). Figure \ref{fig:ellip_time_circu} shows the circulation profiles at different streamwise locations. The curves in (a) and (b) start around $r/R=0.2$ because of the presence of the fairwater at these two locations. An experimental measurement from \cite{jessup-1989} is also shown in (c), which agrees well with the simulation result. Overall, we see that all profiles have very similar shapes and amplitudes, which confirms that the circulation is indeed roughly conserved. Nevertheless, viscous effects are still evident since the local narrow peaks are gradually smoothed out as the flow travels downstream. These curves also signify that the load is mostly concentrated around the mid-section (i.e., $r/R=0.5$) of each blade. The large peak around $r/R=0.9$ in the near filed indicates that the propeller is also heavily loaded around the tips, which is consistent with the strong tip vortices that we have observed in the flow field.

\subsubsection{Pressure field}

\begin{figure}[H]
\centering
\includegraphics[width=5.2in]{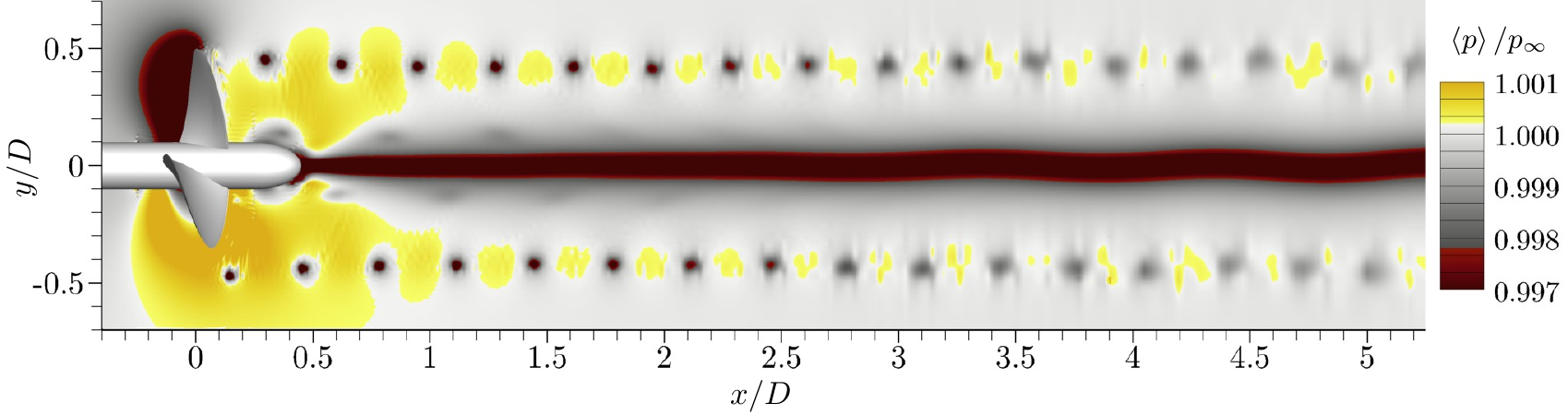}
\caption{Contours of phase-averaged pressure in the central $x$-$y$ plane.}
\label{fig:ellip_phase_p}
\end{figure}

Figure \ref{fig:ellip_phase_p} shows the phase-averaged pressure field in the central $x$-$y$ plane. Comparing with the isosurfaces of Q-criterion in Fig. \ref{fig:ellip_phase_Qcr}, we see that the tip vortices show up as local pressure minima along the edge of the slipstream. Meanwhile, the fairwater wake, especially the hub vortex,  is a very low pressure region, and is responsible for the large drag on the fairwater (see Tab. \ref{tab:ktkq_mean2}).

\begin{figure}[H]
\centering
\includegraphics[width=6.4in]{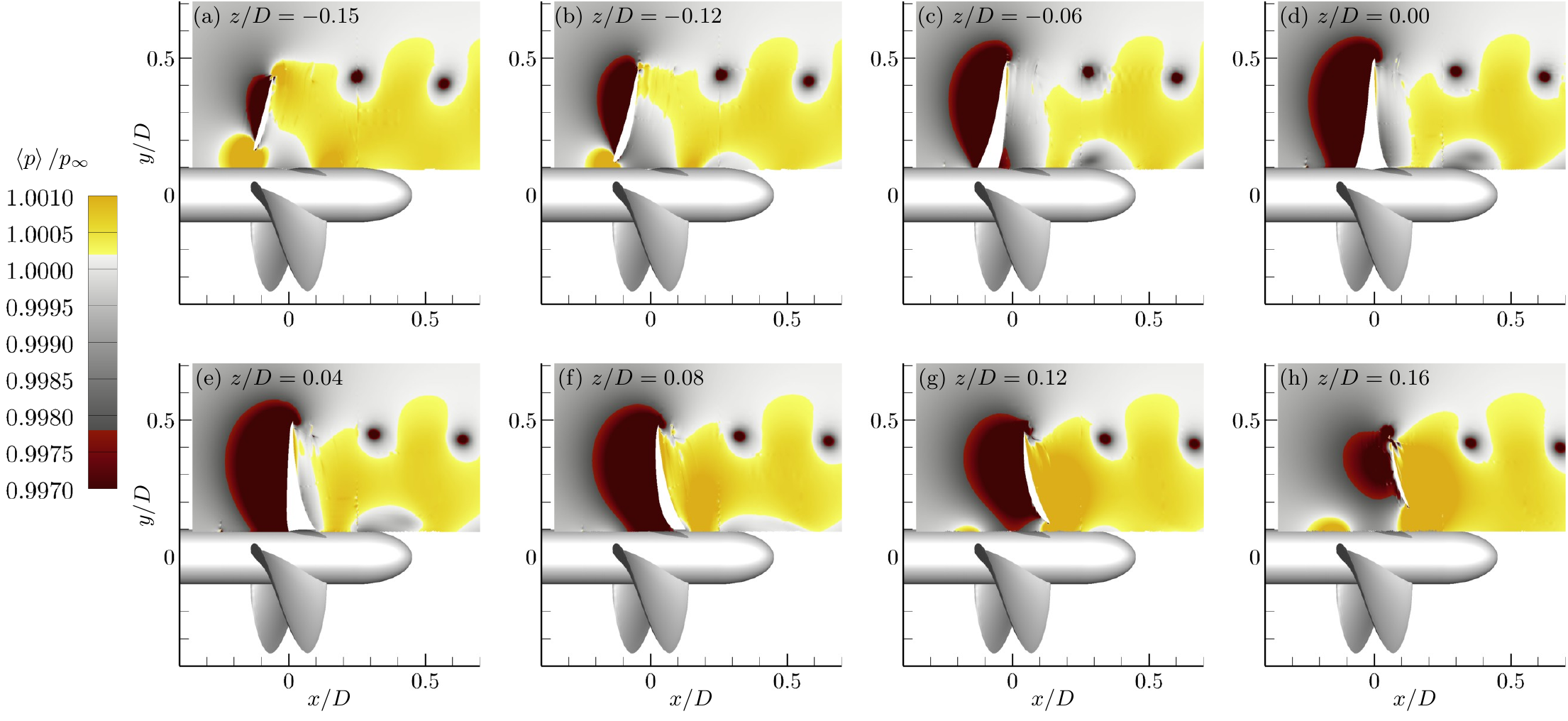}
\caption{Contours of phase-averaged pressure in different $x$-$y$ planes through the top blade.}
\label{fig:ellip_p_xy}
\end{figure}

Of great importance is the pressure distribution around the blades, which directly affects the thrust and torque on the propeller. To see the pressure effects on the thrust more clearly, we have plotted in Fig. \ref{fig:ellip_p_xy} several slices in the $x$-$y$ plane through the top blade at different spanwise (i.e., $z$) locations. From (a) to (h), we are moving from the leading edge to the trailing edge of the blade (refer to Fig. \ref{fig:geo_ellip}) along the $z$ direction. The suction side is on the left, and the pressure side is on the right. As expected, we see that most part of the suction side is in a low pressure region, and the pressure side is in a high pressure region. When we go from the leading edge to the trailing edge, the size of the low pressure region first increases and then decreases, with the maximum size around the mid-span, i.e., $z=0$. In contrast, the size of the high pressure region first decreases, and then increases. These pressure distributions apparently suggest that the thrust load is more concentrated on the trailing portion ($z>0$) of the blade.

\begin{figure}[!hbt]
\centering
\includegraphics[width=6.4in]{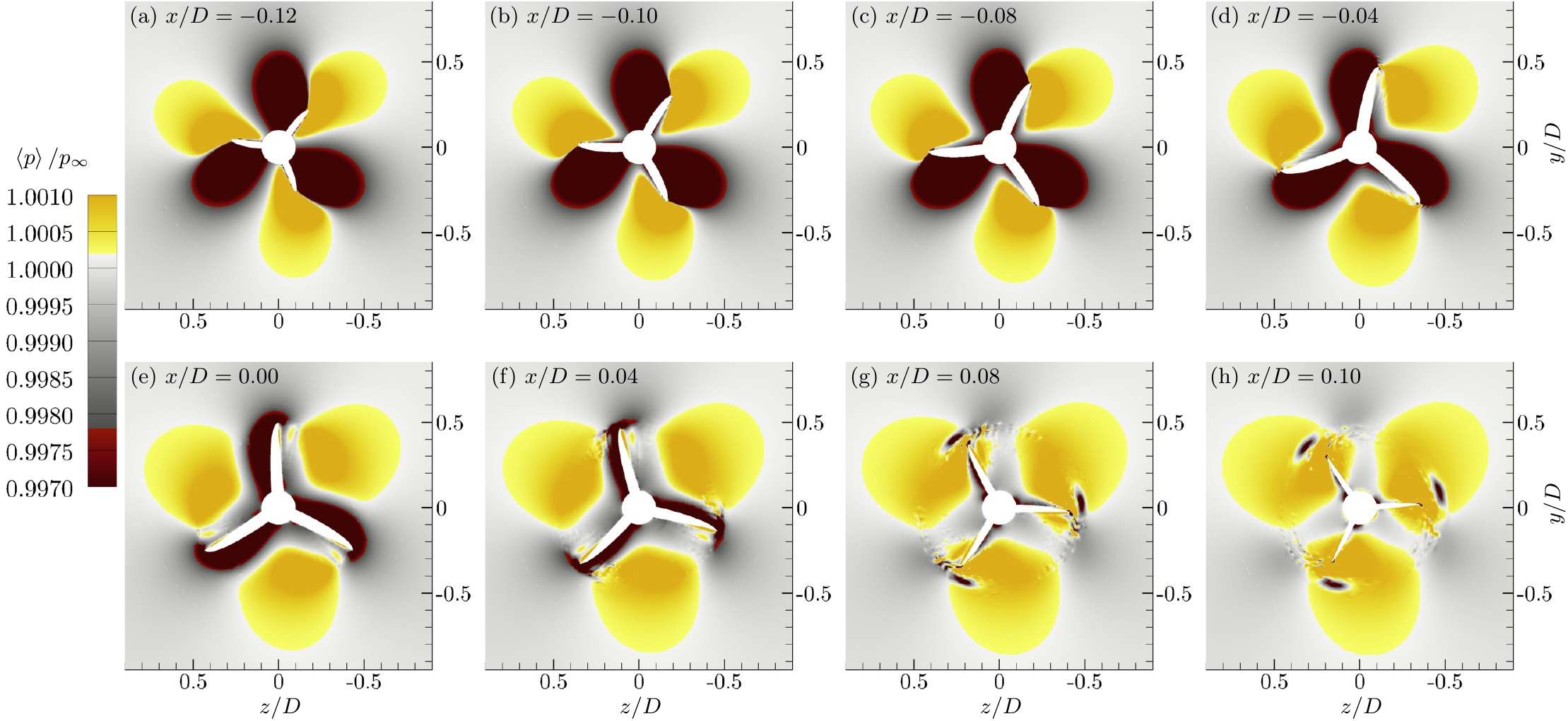}
\caption{Contours of phase-averaged pressure in different $y$-$z$ planes through the blades.}
\label{fig:ellip_p_yz}
\end{figure}

Similarly, the contribution from different parts of a blade to the torque can be visualized through the pressure distribution in different $y$-$z$ planes through the blades as shown in Fig. \ref{fig:ellip_p_yz}. Again, from (a) to (h) we are moving from the leading edge to the trailing edge (refer to Fig. \ref{fig:geo_ellip}), but along the $x$ direction this time. Taking the top blade for example, the suction side is on the left, and the pressure side is on the right. From this perspective, the suction side is almost always in a low pressure region. In (a)-(d), a large portion of the pressure side actually has low surface pressure (although
there is a large high-pressure ``bubble'', but it is detached from the blade surface). In contrast, the pressure side has increased pressure in (e)-(h). This pressure distribution results in higher torque on the trailing portion of the blade. We also notice that the tips of the cross-sections have the largest pressure difference most of the time. Considering the fact that the tips also has the largest arms in the cross-sections, the torque is therefore also very heavily loaded around the edge of each blade.

\begin{figure}[H]
\centering
\includegraphics[width=5.2in]{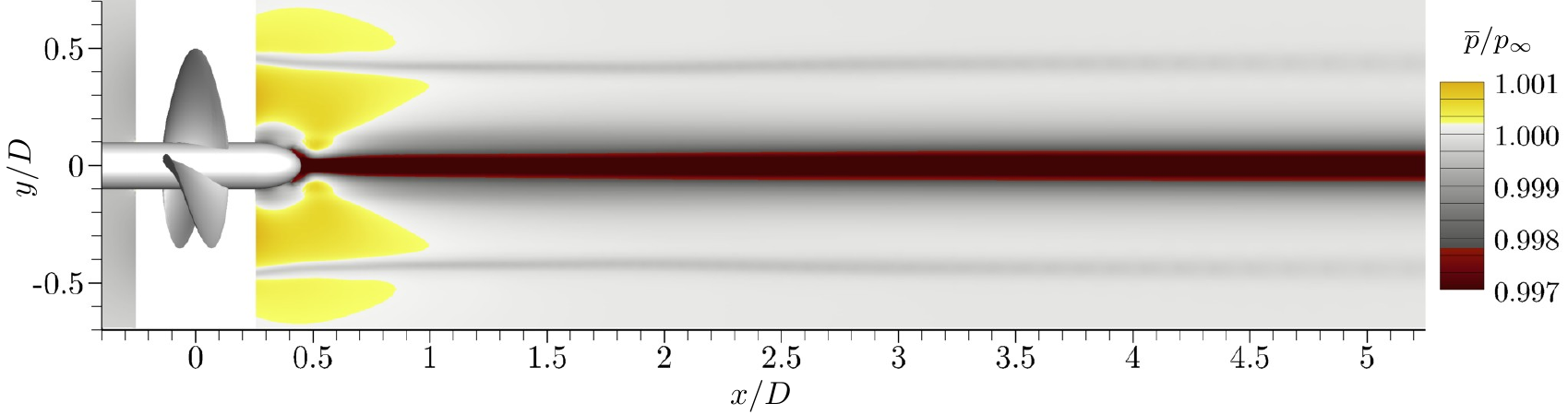}
\caption{Contours of time-averaged pressure in the central $x$-$y$ plane.}
\label{fig:ellip_time_p}
\end{figure}

\begin{figure}[H]
\centering
\includegraphics[width=3.6in]{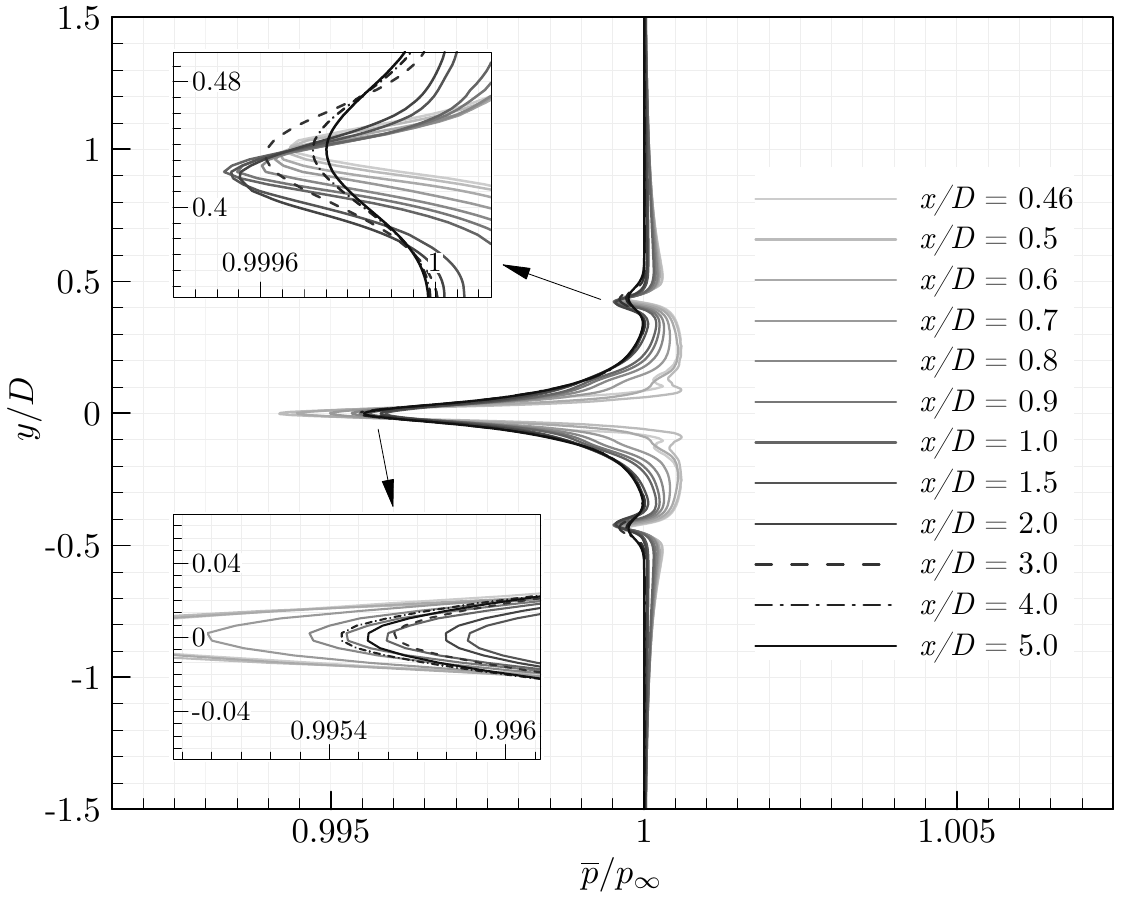}
\caption{Profiles of time-averaged pressure at different streamwise locations in the central $x$-$y$ plane.}
\label{fig:ellip_time_p1d}
\end{figure}

The time-averaged pressure field in the central $x$-$y$ plane is shown in Fig. \ref{fig:ellip_time_p}, and a series of pressure profiles are given in Fig. \ref{fig:ellip_time_p1d}. From the contours, it is seen that on average the propeller generates an obvious high-pressure region in the near field $x/D<1.0$ (inside and outside of the slipstream). Other than this region, the propeller's effects on the pressure field are mostly contained within the slipstream. The hub vortex represents the pressure minima of the whole flow field. From the profiles, we notice three local pressure minima around $r/D=0$, $r/D \approx 0.12$ (only in the very near field), and $r/D\approx 0.42$. They actually correspond to the hub vortex, the root vortices, and the tip vortices, respectively. The pressure recovery in the blade wake ($0.1<r/D<0.5$) is evident as the flow goes downstream. In contrast, we do not see consistent pressure recovery in the fairwater wake ($r/D<0.1$) due to the strong hub vortex.

\subsection{Fairwater effects}
\begin{figure}[H]
\centering
\includegraphics[width=4.2in]{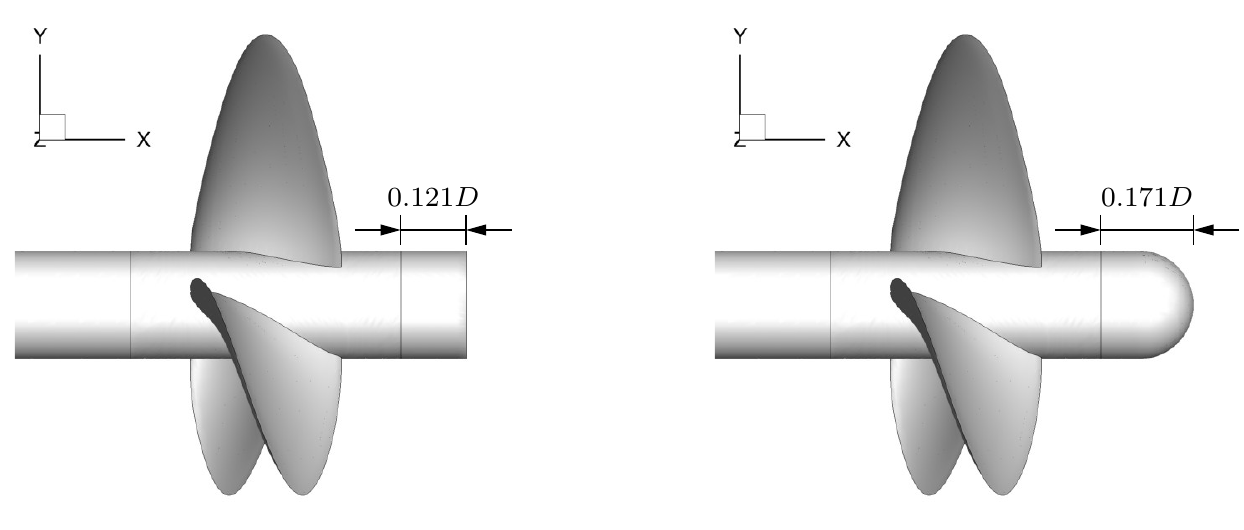}
\caption{DTMB 4119 with cylindrical fairwater (left) and hemispherical fairwater (right).}
\label{fig:fairwater_geo}
\end{figure}

The fairwater is usually not considered in propeller design. A user has the freedom to choose a fairwater based on their preference or the availability of parts. The quantitative effects of fairwater shape have rarely been studied. In this section, we briefly study two more fairwater shapes: cylindrical and hemispherical, to compare with the ellipsoidal one from the previous sections. For a fair comparison, we require the three fairwaters to have the same surface area so that they contact with the same amount of fluids. This means that the fairwaters will have different lengths. The 1:2 ellipsoidal fairwater has a length of $0.2D$ as shown in Fig. \ref{fig:geo_ellip}. The geometries and sizes of the other two are shown in Fig. \ref{fig:fairwater_geo}. As marked in the figure, the cylindrical fairwater has a length of $0.121D$, and the hemispherical one has a length of $0.171D$ (a hemisphere of radius $0.1D$ sitting on top of a cylinder whose height is $0.071D$). Overall, the shape is more elongated (streamlined) as the shape changes from cylindrical to hemispherical and then to ellipsoidal.

The loads for the above two configurations are summarized in Tabs. \ref{fig:ktkq_flat} and \ref{fig:ktkq_hemi}, respectively. The overall loads on the hubs and the torques on the fairwaters are once again negligibly small. Comparing with the ellipsoidal configuration (see Tab. \ref{tab:ktkq_mean2}), we notice that the blades in both configurations here have smaller thrust and torque. However, the efficiencies of the blades (excluding fairwater contributions) in all three configurations stay almost unaffected as shown in the first row of Tab. \ref{tab:efficiency}, where the efficiency differences are around $0.3\%$. The drags on the fairwaters, on the other hand, are dramatically different for the three configurations. The cylindrical fairwater has the largest drag, followed by the hemispherical one, and then the ellipsoidal one. The overall efficiencies (with fairwater contributions included) are also summarized in Tab. \ref{tab:efficiency}. We see that the cylindrical, the hemispherical, and the ellipsoidal fairwaters reduce the propeller's overall efficiency by $3.3\%$, $2.9\%$, and $2.1\%$, respectively. We need to emphasize that these numbers are not small for a propulsion system, and we also need to repeat that these numbers are underestimated based on the assumptions that we made in Sec. \ref{sec:loads}. Since pressure contribution dominates the drag on a fairwater, it is worth looking into the pressure fields to see how the fairwaters affect the pressure distributions.
\begin{table}[H]
\setlength{\tabcolsep}{2mm}
\centering
\begin{tabular}{>{\centering}m{1.4cm} m{1.0cm} >{\centering}m{1.3cm} >{\centering}m{1.3cm} >{\centering}m{1.3cm} >{\centering}m{1.3cm} >{\centering}m{1.3cm} >{\centering\arraybackslash}m{1.3cm}}
\hline
                           &        & $K_T$ & $K_{T,p}$ & $K_{T,v}$ & $K_Q$ & $K_{Q,p}$ & $K_{Q,v}$ \\
\hline
\multirow{2}{*}{blades}    & mean   & ~0.1503 & ~0.1507 & -4.8E-4 & 0.0271 & 0.0268\hphantom{2} & 2.4E-4 \\
                           & r.m.s. & ~1.4E-4 & ~1.4E-4 & ~1.1E-7 & 2.8E-5 & 2.8E-5\hphantom{2} & 5.8E-9 \\
\hline
\multirow{2}{*}{hub}       & mean   & -9.8E-5 & -8.5E-7 & -9.7E-5 & 4.0E-6 & 2.9E-8\hphantom{2} & 3.9E-6 \\
                           & r.m.s. & ~1.1E-7 & ~3.9E-8 & ~9.4E-8 & 1.1E-8 & 2.2E-9\hphantom{2} & 1.0E-8 \\
\hline
\multirow{2}{*}{fairwater} & mean   & -4.9E-3 & -4.9E-3 & -2.9E-5 & 1.5E-6 & 1.3E-11            & 1.5E-6 \\
                           & r.m.s. & ~6.0E-5 & ~6.0E-5 & ~1.2E-7 & 1.9E-8 & 7.9E-12            & 1.9E-8 \\
\hline
\end{tabular}
\caption{Loads on different parts of DTMB 4119 with a cylindrical fairwater.}
\label{fig:ktkq_flat}
\end{table}
\begin{table}[H]
\setlength{\tabcolsep}{2mm}
\centering
\begin{tabular}{>{\centering}m{1.4cm} m{1.0cm} >{\centering}m{1.3cm} >{\centering}m{1.3cm} >{\centering}m{1.3cm} >{\centering}m{1.3cm} >{\centering}m{1.3cm} >{\centering\arraybackslash}m{1.3cm}}
\hline
                           &        & $K_T$ & $K_{T,p}$ & $K_{T,v}$ & $K_Q$ & $K_{Q,p}$ & $K_{Q,v}$ \\
\hline
\multirow{2}{*}{blades}    & mean   & ~0.1502 & ~0.1507 & -4.8E-4 & 0.0271 & 0.0269\hphantom{2} & 2.4E-4 \\
                           & r.m.s. & ~3.2E-4 & ~3.2E-4 & ~5.6E-7 & 6.2E-5 & 6.2E-5\hphantom{2} & 3.0E-8 \\
\hline
\multirow{2}{*}{hub}       & mean   & -9.1E-5 & -9.2E-7 & -9.0E-5 & 3.9E-6 & 2.9E-8\hphantom{2} & 3.9E-6 \\
                           & r.m.s. & ~1.8E-7 & ~5.2E-8 & ~1.6E-7 & 2.9E-8 & 3.4E-9\hphantom{2} & 2.9E-8 \\
\hline
\multirow{2}{*}{fairwater} & mean   & -4.4E-3 & -4.4E-3 & -3.9E-5 & 1.3E-6 & 3.5E-11            & 1.3E-6 \\
                           & r.m.s. & ~1.6E-4 & ~1.6E-4 & ~2.8E-7 & 4.0E-8 & 5.8E-10            & 4.0E-8 \\
\hline
\end{tabular}
\caption{Loads on different parts of DTMB 4119 with a hemispherical fairwater.}
\label{fig:ktkq_hemi}
\end{table}
\begin{table}[H]
\setlength{\tabcolsep}{2mm}
\centering
\begin{tabular}{>{\centering}m{4.5cm} >{\centering}m{2.2cm} >{\centering}m{2.2cm} >{\centering\arraybackslash}m{2.2cm}}
\hline
                              & cylindrical & hemispherical  & ellipsoidal    \\ \hline
blades~ efficiency (excl. FW) & 0.7353      & 0.7348      & 0.7326       \\
overall efficiency (incl. FW) & 0.7113      & 0.7133      & 0.7171       \\
efficiency loss ~~~ (from FW) & -3.3\%      & -2.9\%      & -2.1\%       \\ \hline
\end{tabular}
\caption{Blade efficiency and overall efficiency for different fairwater (FW) configurations.}
\label{tab:efficiency}
\end{table}

\begin{figure}[H]
\centering
\includegraphics[width=5.2in]{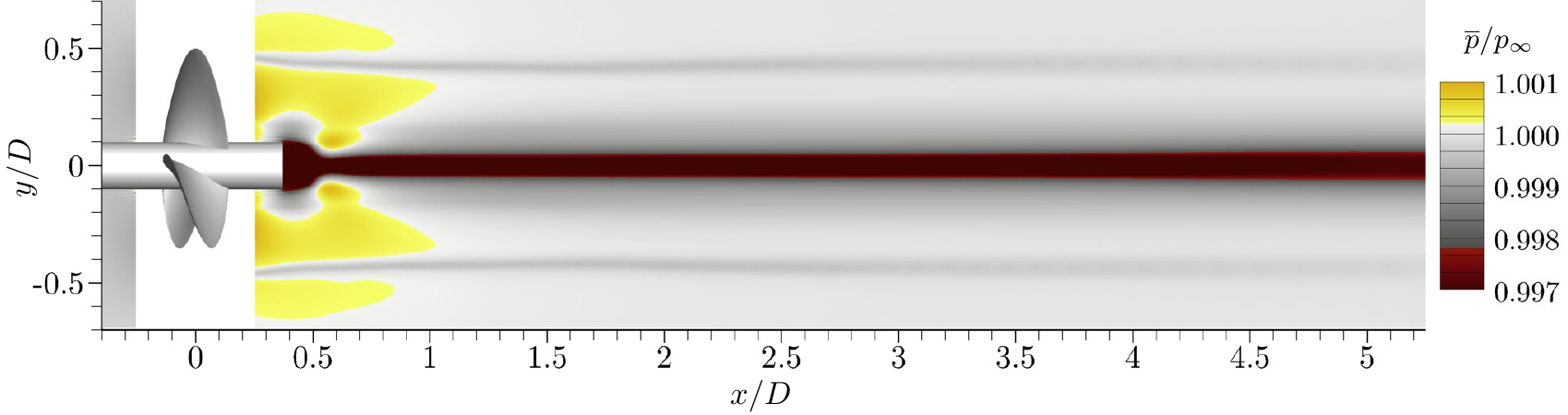}   \\
\includegraphics[width=5.2in]{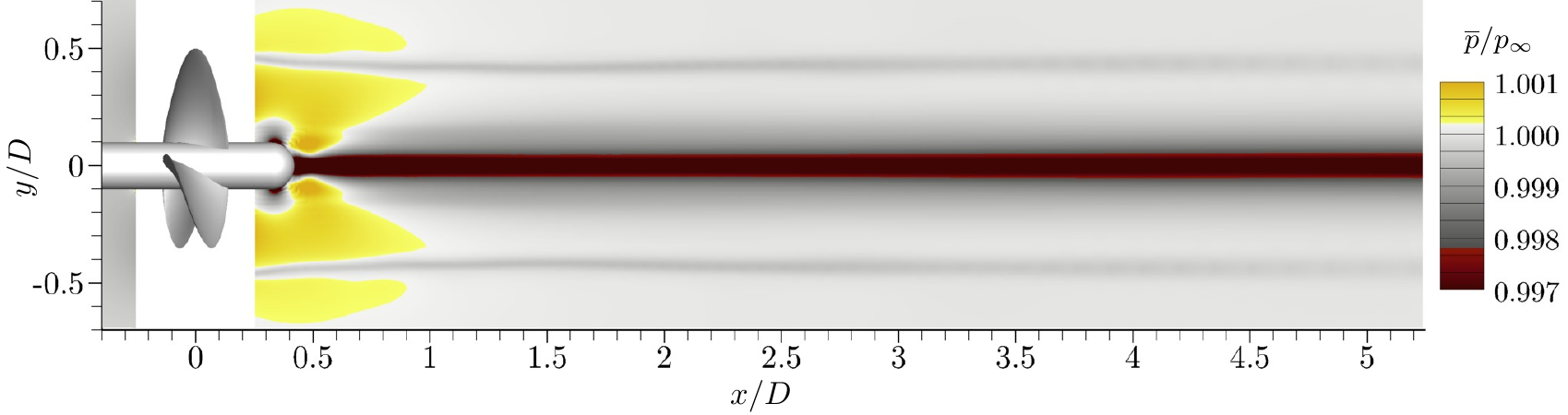}
\caption{Time-averaged pressure in the central $x$-$y$ plane of DTMB 4119 with cylindrical and hemispherical fairwaters.}
\label{fig:flat_hemi_p}
\end{figure}

The time-averaged pressure fields for the above two configurations are shown in Fig. \ref{fig:flat_hemi_p}. Comparing with the ellipsoidal configuration in Fig. \ref{fig:ellip_time_p}, we see that the three configurations overall have very similar pressure distribution in the wake, which explains why the blade efficiencies are not affected much. The differences are mostly limited to the region immediately downstream of the fairwater. The cylindrical fairwater generates a very large low-pressure region at its end. The hemispherical fairwater has a low-pressure region of similar size to that of the ellipsoidal one. However, at the junction of the hemispherical fairwater and the hub, there is another small but very low-pressure region due to the geometric change, which leads to an overall larger drag for this configuration than the ellipsoidal one.

\section{Summary}
\label{sec:summary}

The first high-order eddy-resolving simulation of a marine propeller has been successfully performed in this work using a recently developed sliding-mesh method. This method combines the flux reconstruction framework and a new dynamic curved mortar approach to deal with the complex rotating geometry of a propeller without sacrificing the high-order accuracy at all. Even on a very coarse mesh with less than one-fourth million cells, the method predicts the propeller loads very accurately under a wide range of working conditions, and also captures the flow structures with a lot of details. Moreover, this method allows both phase and time averaging on the same set of grid, and thus can provide more information about a flow field.

Through visualization of vortical flow structures, it is revealed that when the advance coefficient $J$ decreases, the strengths of the major vortices grow and flow instability gradually develops. The instability first comes from tip-tip vortex interaction, then tip-tip as well as tip-hub vortex interactions, and finally the trailing-edge vortices become strong enough and start playing an important rule when $J$ is sufficiently small. At the design condition, the sources of each tip vortex are identified through velocity isosurfaces to be a strand of decelerated flow from the leading edge and a strand of accelerated flow from the trailing edge of each blade.

A comparison between the present high-order simulation and a previous low-order one on the same propeller has clearly demonstrated the low-dissipation advantage of the high-order method, which has allowed accurate prediction of the loads under all working conditions. In contrast, the high-dissipation of the low-order method completely failed the mission for large $J$ that generates weak flow vortices. Detailed load analysis at the design condition has revealed that the major loads are the blade thrust and torque as well as the fairwater drag, and pressure contribution dominates these loads. The pressure field and the circulation distribution show that the blade loads concentrate more on the trailing portion as well as the radial mid-section of each blade.

By studying three fairwaters of different shapes, it is found that these fairwaters do not have obvious effects on the blade performance in the present setups. They, however, do dramatically change the pressure distribution on their surfaces, resulting in different induced drags and different performance degradation to the overall propulsion system. More specifically, we see an efficiency loss of at least $3.3\%$, $2.9\%$, and $2.1\%$, from the cylindrical, the hemispherical, and the ellipsoidal fairwater, respectively. It remains to be investigated whether there is an optimum fairwater shape that can minimize the efficiency loss.

\section*{Acknowledgment}
This work was supported by: the Office of Naval Research (Grant N00014-18-1-2265), the George Washington University, and Clarkson University. The computing resources were provided by the Navy DoD Supercomputing Resource Center through the High Performance Computing Modernization Program. The authors would like to thank Dr. Thad Michael from the Naval Surface Warfare Center Carderock Division for providing the propeller geometry and for helpful discussions. Thanks are also due to Dr. Zihua Qiu for helping on part of the mesh generation.

\section*{Dedication}
This paper is dedicated to Prof. Antony Jameson for celebrating his 85th birthday and his immense and continuing contributions to the many aspects of computational fluid dynamics.


\newpage
\biboptions{sort&compress}
\bibliographystyle{elsarticle}
\bibliography{references}



\end{document}